\documentclass{article}

\usepackage[margin=1in]{geometry}

\usepackage{graphicx} 
\usepackage{amsmath}
\usepackage{amsfonts}
\usepackage{amsthm}
\usepackage{mathtools}
\usepackage{xcolor}
\usepackage{hyperref}
\usepackage{tikz}
\usepackage{pgfplots}
\usepackage{booktabs}
\usepackage{algorithm}%
\usepackage{algorithmicx}%
\usepackage{algpseudocode}%
\usepackage{listings}%
\usepackage{url}
\usepackage[most]{tcolorbox}
\newtcbox{\mymath}[1][]{%
	nobeforeafter, math upper, tcbox raise base,
	enhanced, colframe=blue!30!black,
	colback=blue!30, boxrule=1pt,
	#1}

 \newcommand{\hide}[1]{}

\newtheorem{theorem}{Theorem}
\newtheorem{corollary}{Corollary}
\newtheorem{problem}{Problem}
\newtheorem{formulation}{Formulation}

\def\R{\mathbb{R}}

\def\bx{x}
\def\bs{s}

\newcommand{\com}[1]{\hfill {\color{blue} /\!/ #1}}

 \newcommand{\spara}[1]{\paragraph{#1}}
\newcommand{\fabian}[1]{\textcolor{red}{[Fabian: #1]}}
\newcommand{\labis}[1]{\textcolor{green!80!black}{[Labis: #1]}}

\def\figsp{\vspace{-15pt}}

\begin{document}

\title{Countering Election Sway: \\ Strategic Algorithms in Friedkin-Johnsen Dynamics}

\author{Dragos Ristache \\ dragosr@bu.edu \\ Boston University \and Fabian Spaeh \\ fspaeh@bu.edu \\ Boston University \and  Charalampos E. Tsourakakis \\ ctsourak@bu.edu \\ Boston University}

\maketitle

\begin{abstract}
     Social influence profoundly impacts individual choices and collective behaviors in politics. In this work, driven by the goal of protecting elections from improper influence, we consider the following scenario: an individual, who has vested interests in political party $Y$, is aware through reliable surveys that parties $X$ and $Y$ are likely to get 50.1\% and 49.9\% of the vote, respectively. Could this individual employ   strategies to alter public opinions and consequently invert these polling numbers in favor of party $Y$?

We address this question by employing: (i) the  Friedkin-Johnsen (FJ) opinion dynamics model, which is mathematically sophisticated and effectively captures the way individual biases and social interactions shape opinions, making it crucial for examining social influence, and (ii) interventions similar to those in Asch's experiments, which involve selecting a group of stooges within the network to spread a specific opinion. We mathematically formalize the aforementioned motivation as an optimization framework and establish that it is NP-hard and inapproximable within any constant factor. We introduce three efficient polynomial-time algorithms. The first two utilize a continuous approach: one employs gradient descent with Huber's estimator to approximate the median, and the other uses a sigmoid threshold influence function. The third utilizes a combinatorial greedy algorithm for targeted interventions. Through comparative analysis against various natural baselines and using real-world data, our results demonstrate that in numerous cases a small fraction of nodes chosen as stooges can significantly sway election outcomes under the Friedkin-Johnsen model.

\end{abstract}

\section{Introduction}
\label{sec:intro}
Social media platforms such as Facebook, Twitter, Instagram, and Reddit have created positive social outcomes on major issues such as voter registration and mobilization within authoritarian regimes, but there is increasing concern that they are a major contributor to the rise of political dysfunction seen in the USA and some Western democracies since the early 2010s~\cite{haidt2019dark,mcnamee2020zucked}.   America has a widening political divide, and it stands out above other nations, according to a recent study by Stanford economists \cite{boxell2020cross}.  In 1994,  49\% of Americans held mixed political views with the rest of the mass divided between liberal and conservative \cite{pew2}, while the number of moderates  decreased to 39\% by 2014 \cite{bbc}. This political division is more pronounced than ever; Republicans and Democrats are arguably further apart ideologically than ever  before \cite{nytimes,bbc2,pew1}. For instance, an increasing fraction of Democrats and Republicans view the presidential candidate of the other party ``very unfavorable''. According to Pew Research in 1994 the fraction of  Republicans  that deemed  Democrats unfavorable was 17\% while in the recent presidential elections this fraction was 58\%. Abramowitz  and Webster show that partisan identities have become increasingly associated with divisive issues  in American society, including racial, cultural and ideological issues. This is reflected clearly into the elections' outcomes; the Democratic share of the House vote and the Democratic share of the presidential vote has increased from 0.54 in 1972 to 0.97 by 2018~\cite{abramowitz2015all}. 
 The same polarizing phenomenon is observed on a variety of topics, not just politics, including vaccination and the security measures against COVID-19~\cite{klein2020we,smith2019mapping}.  

Due to the planetary scale of social media, despite the fact that U.S. law bans foreign nationals from making certain expenditures or financial disbursements for the purpose of influencing federal elections, an adversary  has an unprecedented opportunity to shape public opinion using computational propaganda tools {\it remotely}. Few years ago,  the U.S. Department of Justice indicted numerous Russian agents that were  affiliated with the St. Petersburg-based Internet Research Agency (IRA), an organization that allegedly engaged in political and electoral interference operations in the United States which included the purchase of American computer server space, the creation of hundreds of fictitious online personas, and the use of stolen identities of persons from the United States.  According to the indictment issued by Robert Mueller against the Internet Research
Agency, the core strategy was to increase disaffection, distrust, and polarization in American politics~\cite{usjustice}. This can be achieved by using malicious accounts that attack moderate politicians, spread  misinformation in order to bolster controversial candidates, and increase societal polarization~\cite{allcott2017social,rolling}. 

It is not only foreign actors, or malicious entities that manipulate opinion dynamics through  social media. Most notably, Cambridge Analytica used personalized social influence techniques that  exploited personality traits of voters to sway voters away from Hillary Clinton towards Donald Trump~\cite{nix}.  However, there exist plenty of less known ``Cambridge Analytica'' cases. For example, Rally Forge is a private organization for sustainable natural resource conservation. It was proved that it orchestrated astroturfing campaigns, in which a combination of real and inauthentic accounts posted comments and replies to relevant conversational threads to create a misconception that public opinion lied on one side of a particular topic~\cite{astroturf}. In addition to such private organizations whose self-interest lies in manipulating opinion dynamics,    corporations design recommendation systems to maximize user engagement for revenue; however, this can unintentionally lead to increasing societal polarization and radicalization ~\cite{ledwich2019algorithmic}.

In this paper we focus on a problem that is rooted in timely real-world scenarios~\cite{nix,haidt2019dark,mcnamee2020zucked,abramowitz2015all,usjustice}. Specifically, this issue revolves around the unstudied yet crucial concept of the median in equilibrium vectors. To illustrate the importance of this idea, imagine the following scenario: an individual, who has vested interests in political party $Y$, is aware through reliable surveys that parties $X$ and $Y$ are likely to get 50.1\% and 49.9\% of the vote, respectively. Could this individual employ social media strategies to alter public opinions and consequently invert these polling numbers in favor of party $Y$?  Alternatively, can an individual boost the median even if the majority is already voting for their preferred party?

More broadly, what tools are available for a potential manipulator to influence election results? We operate under the assumption that such an individual has comprehensive knowledge of both the network structure and the opinion dynamics described by the Friedkin-Johnsen (FJ) dynamics, c.f.~\cite{friedkin1997social} and   Equation~\eqref{eq:fj}. One objective of our study is to determine whether small interventions have the power to tip an election or, more universally, to shift the median  viewpoint in a given contentious subject towards a particular side.   We focus on interventions similar to those used by Solomon Asch, where certain participants are turned into {\it stooges}. Asch's pioneering experiments, central to social psychology and the study of opinion dynamics, explored how much social pressure from a majority could compel an individual to conform~\cite{asch1955opinions}. Essentially, our objective is to explore the challenge of influencing election results within the Friedkin-Johnsen (FJ) model of opinion formation by strategically selecting a limited number of stooges, similar to Asch's renowned conformity experiments. In the FJ model, this type of intervention is effectively represented by altering the resistance parameter of the stooges~\cite{abebe2020opinion,ristache2024wiser}. For further details, see Section~\ref{sec:proposed}. We state this informally as the next problem.

\begin{tcolorbox}
\begin{problem}[\textsc{Informal Statement - Flipping the Median under the FJ Dynamics}]
\label{prob:informal}
Given a network of agents $G$ whose opinions form according to the FJ model, and an integer $k$, choose a set of $k$ stooges from the node set so that the median of the opinions at equilibrium changes significantly.
\end{problem}
\end{tcolorbox}


\noindent
We make the following contributions
related to this problem.

$\bullet$
We introduce a novel formulation for altering the median of opinions at equilibrium under the FJ dynamics, subject to Asch-like interventions. On one hand, the median presents substantial challenges because it is a non-smooth function of the equilibrium. On the other hand, it effectively models the contemporary issue of influencing elections through adversarial interventions on social media.    
To the best of our knowledge, Problem~\ref{prob:informal}  has not been studied before in any prior works.

$\bullet$
We prove that maximizing the median and any quantile under the FJ model is NP-hard and inapproximable to any constant factor.   

$\bullet$
We develop a continuous optimization framework that  utilizes two different approaches. The first uses Huber's M-estimators to approximate the median effectively. We design a robust gradient descent heuristic and provide a way to choose a good value of Huber's function parameter $c$. This allows us to run our algorithm only once instead of trying it on a large number of possible values.     
We develop an alternative objective via the sigmoid threshold influence function. We also demonstrate that Problem~\ref{prob:elections} can be solved exactly in polynomial time for the
special case of directed trees.

$\bullet$
We demonstrate that the straightforward discrete greedy algorithm is computationally impractical for even small-scale networks, resulting in a complexity of $\Omega(n^4)$, where $n$ is the total number of actors. Consequently, we develop a lazy version of the greedy approach that in practice enables us to apply it to larger networks.
    
$\bullet$
We conduct extensive experiments on real-world datasets and discover that our strategies effectively identify a small number of pivotal agents capable of shifting the median. This insight is vital for understanding how to protect the network against adversarial disturbances in the FJ model.

 \section{Related work} 
\label{sec:rel}

\spara{Opinion Dynamics and the Friedkin-Johnsen Model.} The exchange of opinions   among people is a core social activity that influences almost every social, political, and economic endeavor.  Opinion dynamics  models  have been used in various disciplines to model social learning~\cite{acemoglu2011opinion,acemouglu2013opinion,proskurnikov2017tutorial,proskurnikov2018tutorial,lorenz2007continuous,grabisch2020survey}.   In Friedkin \cite{friedkin2015problem}, it is highlighted that the opinions of individuals represent their cognitive perceptions of certain entities, such as specific issues, events, or other people. This can be seen in manifested attitudes as mentioned by Abelson~\cite{abelson1964mathematical}, Hunter \cite{hunter1984mathematical} or in the personal certainties of belief as noted by Halpern~\cite{halpern1991relationship}. From a mathematical perspective, opinions can be understood as scalar or vector values related to a set of agents.  
There exist several types of opinion dynamics models, see the tutorial by Proskurnikov and Tempo~\cite{proskurnikov2017tutorial,proskurnikov2018tutorial}.
We focus on discrete-time models where the opinions get updated in rounds.
There exist two types of discrete-time models,  models where the opinions are discrete  ~\cite{kempe2003maximizing,richardson2002mining,yildiz2013binary,berenbrink2016bounds} and continuous~\cite{french1956formal,degroot1974reaching,castellano2009statistical,hegselmann2002opinion,friedkin1997social,tsitsiklis1986distributed,lorenz2007continuous}.
The French-DeGroot model describes how individuals reach a consensus through stochastic interactions~\cite{french1956formal,degroot1974reaching}.  Let $G(V,E,w)$ be a weighted graph where the weighted adjacency matrix $W$ is row-stochastic. Each node $u \in V$ has an initial opinion $x_u(0) \in [0,1]$ (or, without loss of generality, in  $[-1,1]$) and let $x(k)=(x_u(k))_{u \in V}$ be the vector of opinions in round $k$. The French-DeGroot model updates the opinion vector according to $x(k+1)=W x(k), k=0,1,\ldots$. The convergence criteria are well understood~\cite{proskurnikov2017tutorial}.

Friedkin and Johnsen extended the French-DeGroot model to incorporate individuals' intrinsic beliefs and prejudices~\cite{friedkin1990social}. 
Each node $u\in [n]$ corresponds to a person who has
an \emph{innate opinion} $s_u$ and an \emph{expressed opinion}.
For each node~$u$, the innate opinion $s_u\in[0,1]$ is fixed over time and kept
private; the expressed opinion $x_u(t) \in[0,1]$ is publicly known 
and it changes over time $t\in\mathbb{N}$ due to peer pressure. 
Initially, $x_u(0)=s_u$ for all users $u\in V$. 
At each time
$t>0$, all users $u\in V$ update their expressed opinion $x_u(t)$ as the
weighted average of their innate opinion and the expressed opinions of their
neighbors, as 
follows:
\begin{align}
\label{eq:update-opinions}
	x_u(t) \!
	=\! \dfrac{s_u + \sum_{v\in N(u)} w_{uv} x_v(t)}{1 + \sum_{v\in N(u)} w_{uv}} \! =\! \dfrac{s_u + \sum_{v\in N(u)} w_{uv} x_v(t)}{1 + \deg(u)}
\end{align}
We refer to this model as  the classic FJ model~\cite{friedkin1990social}. In the limit $t\to \infty$, the expressed opinions under mild conditions that are well understood reach an equilibrium $x^\star=(I+L)^{-1}s$.
In this work, we concentrate on the Friedkin-Johnsen model along with an enhanced variant that  delves into the practical consideration  that agents  exhibit varying degrees of susceptibility to persuasion~\cite{cialdini2001science}:
\begin{equation}
\label{eq:dynamics}
\boxed{\textstyle x_u(t+1) = \alpha_u s_u + \frac{1-\alpha_u}{\deg(u)} \sum_{v \in N(u)}  w_{uv} x_v(t) \ \text{~for all~~}u \in V.}
\end{equation}
We refer to this variant as the generalized FJ model~\cite{abebe2020opinion,ghaderi2013opinion,bindel2015bad}, as it is clear that Equation~\eqref{eq:update-opinions} is a special case of Equation~\eqref{eq:dynamics}.  Let us represent the row-stochastic normalized adjacency matrix as \( W \) and the diagonal susceptibility matrix as \( A = {\sf Diag}(\alpha_1, \ldots, \alpha_n) \). With these definitions, we can rewrite Equation~\eqref{eq:dynamics} in  matrix form:
\begin{equation}
\label{eq:fj}
x(t+1) = A s + (I-A) W x(t) 
\end{equation}
By setting $x(t+1)$ equal to $x(t)$, it becomes evident that the generalized FJ equilibrium~\cite{abebe2020opinion,ghaderi2013opinion,bindel2015bad} vector is
\begin{equation}
\label{eq:fj2}
\boxed{x^\star = (I - (I-A) W)^{-1}As} 
\end{equation}
The criteria for convergence have been thoroughly discussed in references such as~\cite{ghaderi2014opinion,proskurnikov2017tutorial}. The FJ model     strikes an excellent balance between being both mathematically rigorous and manageable as a model, while also capturing the realistic dynamics of opinion evolution. Friedkin and Bullo~\cite{friedkin2017truth}  emphasized that the FJ model stands alone as the sole framework that has been subjected to a consistent series of human-subject experiments, enabling the assessment of its predictive capabilities concerning changes in opinions~\cite{friedkin2011social,friedkin2016theory}. It is essential to highlight that $x^\star=x^\star(\mathbf{\alpha},W,s)$ depends on the resistance values $\mathbf{\alpha}$, the topology of the graph $W$, and the innate opinions $\mathbf{s}$.   Our paper examines interventions at the resistance parameter level, \(\mathbf{\alpha}\), which represent an Asch-type intervention involving the recruitment of stooges.

\spara{Optimizing objectives within the Friedkin-Johnsen model} While influence maximization has a well-established history in discrete models, starting with the seminal work of Kempe, Kleinberg, and Tardos~\cite{kempe2003maximizing}, its application in continuous opinion dynamics models has been largely neglected, with most focus on the development of the models themselves. The research initiated by Gionis, Terzi, and Tsaparas~\cite{gionis2013opinion} marked a significant shift by addressing the optimization of the aggregate sum of opinions at equilibrium through the selection of $k$ nodes and consistently fixing their expressed opinion to 1. This approach utilizes the standard Friedkin-Johnsen (FJ) model~\cite{friedkin1990social}, incorporating the concept of stubbornness by keeping the opinions of the selected $k$ nodes constant. 
Musco, Musco, and Tsourakakis~\cite{musco18} further expanded on this by optimizing an objective that balances disagreement and polarization at equilibrium.  Since then, various other formulations and algorithmic solutions have been proposed~\cite{biondi2023dynamics,zhu2022nearly,sun2023opinion,tang2021susceptible,zhu2021minimizing}.  For instance, Gaitonde, Kleinberg, and Tardos~\cite{gaitonde2020adversarial}, along with Chen and Rácz et al.~\cite{chen22,DBLP:journals/corr/abs-2206-08996}, investigated budgeted adversarial interventions on inherent opinions, revealing significant links between spectral graph theory and opinion dynamics. The work of Abebe et al.~\cite{abebe2020opinion}, which is closely related to our research, pioneered adversarial interventions at the susceptibility level using the generalized FJ model. They addressed an unbudgeted optimization problem that is neither convex nor concave, and developed a local search algorithm uncovering a remarkable structure in the local optima of the objective functions. Furthermore, Chan and Shan~\cite{chan2021hardness} demonstrated the NP-hardness of this problem when subject to budget constraints. Closest to our work lies the recent work by Ristache, Spaeh, and Tsourakakis~\cite{ristache2024wiser} that explores the impact of interventions targeting susceptibility on the wisdom of crowds, specifically analyzing how these affect mean squared error (MSE) and polarization, and demonstrating their significant interrelation. It should be noted that the median is a non-smooth statistic that has not been explored in this research area before, despite its significant implications in the context of elections.

\section{Proposed methods} 
\label{sec:proposed}

Our work focuses on the {\it median} of the equilibrium
opinion vector (or simply, the median opinion),
which we defined in Equation~\eqref{eq:fj2}.
The median opinion indicates exactly where
the majority lies, and therefore directly
relates to winning or loosing an election.
In this section, we introduce different
strategies that can be used to optimize
the median in a specific direction.
For simplicity, we only
formally provide an objective with the goal to
maximize the median, but the minimization
objective can be defined analogously.
We introduce the following optimization problem
where we want to maximize the increase in the
median opinion under a fixed budget $k$ of the resistance parameter change. 
For any agent $u \in V$, we may change
its resistance $\alpha_u$, but not its
innate opinion $s_u$; this reflects a simple
attack on someone's susceptibility to persuasion
\cite{abebe2020opinion}.
We measure distances between the original
resistance vector $\alpha$ and modified
resistances $\alpha'$ in an arbitrary
$\ell_p$ norm $\| \cdot \|_p$, and we obtain
one optimization problem for each value
of $p \ge 0$.

\begin{tcolorbox}
\begin{problem}[\textsc{Maximizing the Median under FJ Dynamics}]
\label{prob:max-median} Let $(\alpha, W, s)$ be a network
of generalized FJ dynamics and $k$ be a budget on the change of resistance values. Maximize the objective
$\mathrm{Median}(x^\star(\alpha', W, s))$
where $\alpha' \in [0,1]^n$ is a vector of resistances
such that $\|\alpha' - \alpha\|_p \le k$

\end{problem}
\end{tcolorbox}

We also consider the dual of this problem, which
addresses the question of how many stooges are
needed in order to flip the median beyond
a certain threshold $\theta$. This problem directly addresses the question of what budget is necessary in order to win an election. 

\begin{tcolorbox}
\begin{problem}[\textsc{Flipping the Median under FJ Dynamics}]
\label{prob:elections}
Let $(\alpha, W, s)$ be a network of generalized FJ dynamics
and $\theta \in \R$ a threshold.
Minimize $\|\alpha' - \alpha\|_p$ such that
$\alpha' \in [0,1]^n$ is a vector of resistances
with $\mathrm{Median}(x^\star(\alpha, W, s)) > \theta$.
\end{problem}
\end{tcolorbox}

The two norms of interest in this work are for $p\in\{0,1\}$. The $\ell_0$
pseudo-norm is defined as
$\| \alpha - \alpha' \|_0 = |\{ u \in V : \alpha_u \not= \alpha'_u \}|$
which counts the number of distinct elements.
This corresponds to targeting
$k$ nodes and converting them into stooges. The variant with $p=1$ has also received attention in the literature~\cite{chan2021hardness}.

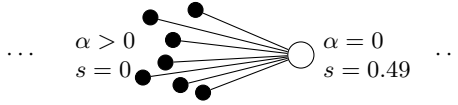
\begin{figure}
    \centering
    \small
    \begin{tikzpicture}
        \node at (-1.7, -0.5) {$\cdots$};
        \node at (4, -0.5) {$\cdots$};
        \node[align=left] at (-0.6, -0.5) {$\alpha>0$\\ $s=0$};
        \node[inner sep=2pt, draw=black, circle, fill] (u1) at (0, 0) {};
        \node[inner sep=2pt, draw=black, circle, fill] (u2) at (0.6, 0.1) {};
        \node[inner sep=2pt, draw=black, circle, fill] (u3) at (-0.1, -0.8) {};
        \node[inner sep=2pt, draw=black, circle, fill] (u4) at (0.3, -0.3) {};
        \node[inner sep=2pt, draw=black, circle, fill] (u5) at (0.2, -0.6) {};
        \node[inner sep=2pt, draw=black, circle, fill] (u6) at (0.4, -0.9) {};
        \node[inner sep=2pt, draw=black, circle, fill] (u7) at (0.7, -1) {};
        \node[inner sep=3.5pt, draw=black, circle, label={[align=left]right:$\alpha=0$\\ $s=0.49$}] (v) at (2, -0.5) {};
        \draw (u1) to (v);
        \draw (u2) to (v);
        \draw (u3) to (v);
        \draw (u4) to (v);
        \draw (u5) to (v);
        \draw (u6) to (v);
        \draw (u7) to (v);
    \end{tikzpicture}
    \caption{An example showcasing challenges in
    Problems~\ref{prob:max-median}
    and \ref{prob:elections}. We show
    an isolated component in a larger
    graph, where the black vertices
    on the left have innate innate opinion
    $0$, but are only connected to the
    white vertex on the right with innate
    opinion $0.5$ but resistance $0$.}
    \label{fig:example}
\end{figure}

We can trivially convert an algorithm for
Problem~\ref{prob:max-median} into an
algorithm addressing Problem~\ref{prob:elections}
by a binary search over the budgets.
A natural heuristic for these optimization
problems is the greedy algorithm, which
in each iteration designates a new node
as a stooge, and the selection is made 
such that it maximizes the increase in the
median opinion.
This solves both problems at the same time,
as we can keep adding stooges until we
flip the median opinion across the
threshold $\theta$.
However, Figure~\ref{fig:example} exemplifies
a difficulty with this approach:
Initially, it may seem that using the
the white node on the right as a stooge
and changing its resistance to $\alpha=1$
results in a big increase in the
median opinions. However,
if the objective is to move the
median beyond a threshold of $\theta=0.5$,
this modification is in vain, since it
does not move any opinion past the
threshold. Furthermore, employing
more stooges in the shown isolated
component does not help with our objective either.
Greedily selecting stooges is therefore
problematic. Other problems with
the greedy algorithm are its inefficiency
on large graphs, since we need to identify
the best node to become a stooge in
each iteration. We will later
discuss the greedy algorithm and how to
apply it on large networks.
We show formally that our targeted
problems are hard to approximate in
Theorem~\ref{thm:hardness}.
In the remainder of this section,
we analyze the hardness and discuss different heuristics
to address Problems~\ref{prob:max-median}
and Problem~\ref{prob:elections}.

\subsection{Hardness of Problem~\ref{prob:max-median}}

Our first key result is that Problem~\ref{prob:max-median} is computationally hard. Specifically, we establish the following inapproximability result. 

\begin{theorem}
    \label{thm:hardness}
    Problem~\ref{prob:max-median}
    for $p=0$
    is inapproximable
    to any multiplicative factor
    unless $\mathsf{P} = \mathsf{NP}$.
\end{theorem}

The proof is based on a novel reduction from the set cover problem~\cite{williamson2011design}, which we defer to the appendix. Furthermore, a straight-forward corollary of our reduction shows that this result carries over not only to the $\frac{1}{2}$-percentile which is the median, but to any $q$-quantile for $0 < q < 1$.

\begin{corollary}
\label{cor:quantile}
Let $(\alpha, W, s)$ be a network of generalized FJ  dynamics,
let $k$ be a budget and $0 < q < 1$  be any fixed quantile value.
The problem of maximizing the $q$-th quantile of $x^\star(\alpha', W, s)$ subject to $\|\alpha' - \alpha\|_0 \le k$ (akin to Problem~\ref{prob:max-median} for $q=\frac{1}{2}$) cannot be approximated to any multiplicative factor unless $\mathsf{P} = \mathsf{NP}$.
\end{corollary}

On the other hand,
for the special case of directed trees,
we can solve this problem optimally
in polynomial time using dynamic programming,
which we show in the appendix.

\subsection{Huber M-Estimator}

To circumvent the hardness of
maximizing the median
and avoid the aforementioned
issues with the greedy algorithm,
we want to turn our attention on
a continuous approximation of the median which
makes our objectives amenable to iterative methods.
The median as a function represents a piecewise-defined
function characterized by inherent discontinuities.
This non-differentiable nature precludes the applicability
of gradient-based optimization techniques, which
fundamentally rely on smooth and continuous objectives
to ensure convergence towards a (potentially locally)
optimal solution. 
Continuous approaches do not commit
to stooges, which could lead to problems as the one
shown in Figure~\ref{fig:example}.
Our approach for tackling Problem~\ref{prob:elections} is
formally stated in Algorithm~\ref{alg:huber-gd}.
However, before delving into details, we
want to motivate our continuous relaxation. We start by
using the Huber M-estimator to derive smooth,
differentiable formulations for the median,
akin to how the softmax function serves as a smooth
approximation for the maximum value
\cite{brown2001smoothed,hampel2011smoothing}.
The Huber M-estimator is a robust statistical estimator that
combines properties of both the sample mean and the sample
median to produce an estimate that is resistant to outliers.
The key idea is to minimize a function of residuals that is
quadratic for small residuals but linear for large residuals.
The Huber loss is parameterized by a tuning constant $c$ and is defined as 
\[
    H_c(x) =
    \begin{cases} 
        \frac 1 2 x^2 & \textrm{if } |x| \le c \\
        c \cdot \left(|x| - \frac 1 2 c\right) & \textrm{otherwise} .
    \end{cases} 
\]
The tuning constant $c$ determines the point at which the loss switches from being quadratic to linear. The intuition behind Huber's M-estimator being a good proxy for the median lies in its loss function, which is less sensitive to outliers than the mean. The mean averages all values equally, giving high influence to outliers, while the median is resistant to outliers but does not take into account ``how far'' the other points are from it.  To find Huber's M-estimator $\hat y$ for a data set \( \{ x_1, x_2, \ldots, x_n \} \), we
solve the minimization
\begin{align}
    \label{eq:y-star}
    \hat y = \hat y_c = \min_y \sum_{i=1}^{n} H_c(x_i - y) .
\end{align}
The extreme values of $c$
give us either the median or
the mean, as $\hat y_0 = \mathrm{Median}(\bx)$
and $\hat y_\infty = \mathrm{Mean}(\bx)$. After introducing the robust and continuous
approximation to the median, we now state a continuous adaption of
Problem~\ref{prob:elections} with $\ell_1$ budget constraint~\cite{chan2021hardness}

\begin{formulation}[\textsc{Minimal Budget for Election Influence via FJ Dynamics}]
\label{prob:elections_formal} 
Consider a network of FJ dynamics represented by $(\alpha, W, s)$, where $c \ge 0$ is the Huber loss parameter, and $\theta \in \R$ represents a threshold.  Minimize $\|\alpha' - \alpha\|_1$, subject to the condition that $\alpha' \in [0, 1]^n$ is a vector of resistances, and $\hat y(\alpha') \ge \theta$, where $\hat y(\alpha')$ is defined as $\min_y \sum_{i=1}^n H_c(x^\star_i(\alpha', W, s) - y)$.

\end{formulation}

\paragraph{Gradient Formulation}
We now show how to derive a gradient
for the Huber's M-estimator $\hat y$.
Since the M-Estimator itself is the
result of optimization problem in
Equation~\ref{eq:y-star},
we first need to analyze properties
of the minimizer.
To this end,
we define the set
$I = \{i : |x^\star_i - \hat y| < c\}$.
For any $y \in \mathbb R$,
\[
    \sum_{i} H_c(x^\star_{i} - y) =
    c \sum_{i \notin I} \left(|x^\star_i - y| - \frac 1 2 c\right)
    + \frac 1 2 \sum_{i \in I} (x^\star_i - y)^2
\]
and we thus have the identity
\begin{multline*}
    0 = \frac{\partial}{\partial \hat y} \sum_i H_c(x^\star_i - \hat y)
    = -c \sum_{i \notin I} \mathrm{sgn}(x^\star_i - \hat y)
    - \sum_{i \in I} (x^\star_i - \hat y) \\
    \iff
    \hat y = \frac 1 {|I|} \Big(
      c \sum_{i \notin I} \mathrm{sgn}(x^\star_i - \hat y) +
      \sum_{i \in I} x^\star_i \Big) .
\end{multline*}
Assuming $x^\star_i \not= \hat y$ for all $i$, we thus obtain
\[
    \frac{\partial \hat y}{\partial x^\star_i} =
    \frac 1 {|I|} 1_{[i \in I]}
\]
where we denote with $1_{[\mathrm{cond}(u)]} \in \R^n$
the vector which is $1$ in the rows corresponding
to $u$ where $\mathrm{cond}(u)$ is true and $0$, otherwise.
It remains to compute the derivative          \def\transMatrix{W}
$ \partial \bx^\star / \partial \alpha $
where $\bx^\star = (I - (I - A) \transMatrix)^+ A \bs$ are the equilibrium opinions and
$A = \mathrm{Diag}(\alpha)$.
Let
\begin{align*}
    X =
    I - (I - A) \transMatrix
    \quad\textrm{and}\quad
    \bx^\star =
    X^+ A \bs .
\end{align*}
Note that we can obtain $\bx^\star$ without
computing the full pseudoinverse $X^+$
by solving the least-squares problem
$\min_\bx \| X \bx - A \bs \|_2$.
We want to show that
\[
    \frac{\partial \bx^\star}{\partial \alpha} =
    X^+ \mathrm{Diag}(\bs) - X^+ \mathrm{Diag}(\transMatrix \bx^\star) .
\]
By the product rule,
\[
    \frac{\partial \bx^\star}{\partial \alpha} =
    \frac{\partial X^+ A \bs}{\partial \alpha} =
    X^+ \otimes \frac{A \bs}{\partial \alpha} +
    \frac{\partial X^+}{\partial \alpha} \otimes A \bs .
\]
We evaluate both terms. The first term is easy:
\[
    X^+ \otimes \frac{\partial A \bs}{\partial \alpha} =
    X^+ \otimes \mathrm{Diag}(\bs) =
    X^+ \mathrm{Diag}(\bs) .
\]
For the second term, we apply the chain rule:
\[
    \frac{\partial X^+}{\partial \alpha} =
    \frac{\partial X^+}{\partial X} \otimes
    \frac{\partial X}{\partial \alpha} =
    \frac{\partial X^+}{\partial X} \otimes
    \frac{\partial A \transMatrix}{\partial \alpha} .
\]
We  use that
$\partial X^+_{ij} / \partial L_{ab} = -X^+_{ia} X^+_{bj}$
since $X$ has full rank.
Furthermore, 
\[
    \frac{\partial (AM)_{ab}}{\partial a_k}
    = 1_{[a = k]} M_{ab} .
\]
Thus,
\begin{align*}
    \left( \frac{\partial X^+}{\partial X} \otimes
    \frac{\partial A \transMatrix}{\partial \alpha} \right)_{ijk}
    &= - \sum_{a,b} X^+_{ia} X^+_{bj} 1_{[a = k]} M_{ab} \\
    &= - \sum_{b} X^+_{ik} X^+_{bj} M_{kb}
    = - X^+_{ik} (\transMatrix X^+)_{kj}
\end{align*}
and
\begin{align*}
    \left( \frac{\partial X^+}{\partial \alpha} \otimes A \bs \right)_{ik} &=
    - \sum_j X^+_{ik} (\transMatrix X^+)_{kj} (A \bs)_j \\ &=
    - X^+_{ik} (\transMatrix X^+ A \bs)_{k} =
    - X^+_{ik} (\transMatrix x^\star)_{k}
\end{align*}
which means
$\partial X^+ / \partial \alpha \otimes A \bs = - X^+ \mathrm{Diag}(\transMatrix \bx^\star)$.
Putting everything together, we get
\begin{align}
    \label{eq:2}
    \nabla \hat y &= \frac 1 {|I|}
        \mathrm{Diag}(\bs - \transMatrix \bx^\star) (X^+)^\top 1_{[i \in I]}
\end{align}
which we can again solve efficiently by solving
the last squares problem
$\min_{z} \| X^\top z - 1_{[i \in I]} \|_2$.
Then, assuming $z$ is the solution
to the least squares problem,
we have that
$\nabla \hat y = \frac 1 {|I|} \mathrm{Diag}(\bs - \transMatrix \bx) z$.

\begin{algorithm}
\caption{Optimizing $\hat y$ with Gradient Ascent}
\label{alg:huber-gd}
\begin{algorithmic}[1]
\Function{Projected Huber}{$W, \alpha_0, s, k, \eta$}
    \State $\alpha \gets \alpha_0$
    \While{not converged}
        \State Let $X = I - (I - A) \transMatrix$ where $A = \mathrm{Diag}(\alpha)$
        \State Solve $\bx^\star = \min_{\bx} \| X \bx - A \bs \|_2$
        \com{Calculate opinions}
        \State Let $\hat y = \min_{y} \sum_i H_c (x^\star_i - y)$
        \com{Huber M-estimator}
        \State Let $I = \{ i : |x^\star_i - \hat y| < c \}$
        \State Solve $\hat z = \min_{z} \|X^\top z - 1_{[i \in I]}\|_2$
        \State Let $\nabla \hat y = \frac 1 {|I|} \mathrm{Diag}(\bs - \transMatrix \bx) \hat z$
        \com{Determine gradient}
        \State $\alpha' \gets \alpha + \eta \nabla \hat y$
        \com{Gradient update}
        \State $\alpha \gets \min \{ \|\alpha - \alpha'\|_2 :
            \|\alpha - \alpha_0\|_1 \le k \}$
        \com{Projection}
    \EndWhile
    \State \Return $\alpha$
\EndFunction
\end{algorithmic}
\end{algorithm}

\spara{Algorithm and complexity}
We optimize the robust median $\hat y$
via gradient ascent with step size
$\eta > 0$, as shown in
Algorithm~\ref{alg:huber-gd}.
For a single gradient computation, we need to
solve the two least squares problems to obtain
$\bx$ via Equation~\eqref{eq:fj2} and
$\nabla \hat y$ via Equation~\ref{eq:2}.
We further carry out a constant number of
matrix multiplications, where each matrix
is of size $n$. In total, a single gradient
computation thus has worst-case running
time $O(n^3)$.

\spara{Selecting the Huber tuning constant $c$}
Our continuous optimization via the Huber
M-Estimator relies on a good choice of
$c$. Generally, this choice is instance-specific
and involves a trade-off between robustness
and approximation quality of the median.
Choosing a value of $c$
that is too small may introduce issues
with vanishing gradients.
We will later give an approach to set an
appropriate value of $c$ in our experiments
in Section~\ref{sec:exp}.

\subsection{Sigmoid Threshold Influence Method}
\label{sec:sigmoid}

We consider a second natural method
for the continuous objective.
However, this method is only suited
to flip the median, as defined in
Problem~\ref{prob:elections_formal}.
Here, we create an objective
that rewards opinions above $\theta$
and penalizes opinions below $\theta$.
Imagine an objective that is simply linear
in the number of nodes $u$ with opinion
$x^\star_u > \theta$.
If we were able to maximize this objective
optimally, we could easily tell whether
the given budget allows us to flip
the median. However, we want 
a continuous reward function to make this
amenable to iterative methods. A
natural choice is the sigmoid function centered
around $\theta$, which is defined
for some temperature $\tau > 0$
as the function
$\mathrm{sigmoid}(x) = 1 / (1 + e^{\tau (\theta-x)})$.
Our objective is
$$
\boxed{     f_{\mathrm{sigmoid}}(\bx^\star) =
        \sum_{u \in V} \mathrm{sigmoid} (x^\star_u) .}
$$
We optimize this objective via
projected gradient ascent on
the resistances $\alpha$. Due to space constraints, we provide the detailed pseudocode
in Algorithm~\ref{alg:sigmoid}
in the appendix.  Note that this objective is only
useful when our sole intent is to flip the
median. If the given budget is not
sufficient to flip the median, we are
not guaranteed to make progress with
the median opinion towards the
threshold $\theta$. For brevity, we refer to this method as $\textsf{Sigmoid}$ in our experiments.

\subsection{Discrete Lazy Greedy}
\label{sec:lazy-greedy}

In this section, we explore the greedy algorithm and its application to Problems~\ref{prob:max-median} and Problem~\ref{prob:elections}, with pseudocode provided in Algorithm~\ref{alg:lazy-greedy}. The algorithm functions by selecting elements from the set of pairs \((u, r) \in V \times \{0, 1\}\), where choosing a pair \((u, r)\) adjusts the resistance of node \(u\) to \(r\). In each iteration, the algorithm selects the pair \((u, r) \in V \times \{0, 1\}\) that maximally increases the median opinion. This process is repeated until either the stooge budget is depleted or the median surpasses the target threshold \(\theta\). This method efficiently addresses both Problem~\ref{prob:max-median} and \ref{prob:elections}, offering solutions applicable across various budget scenarios.

\begin{algorithm}
\caption{Greedily Selecting $k$ Stooges with Laziness $\phi$}
\label{alg:lazy-greedy}
\begin{algorithmic}[1]
\Function{Lazy Greedy}{$W, \alpha, s, k, \phi$}
    \State $m[u, r] \gets \infty$
        for all $u\in V, r\in \{0,1\}$
    \For{i = 1, 2, \dots, k}
        \State $\hat m \gets 0$
        \State $(\hat u, \hat r) \gets (\bot, \bot)$
        \State $\bx \gets \bx^*(G, \alpha, s)$
        \For{$(u, r) \in V \times \{0, 1\}$
            descending in $m[u, r]$}
            \If{$\phi \cdot \hat m \ge \hat m[u, r]$}
                \State {\bf break}
                \com{Abort early}
            \EndIf
            \State Let $\alpha_v^\ddagger = \alpha_v$ for all $u\not=v$ and
            set $\alpha_u^\ddagger = r$
            \State $\bx^{\ddagger} \gets \bx^\star(G, \alpha^{\ddagger}, s)$
            \State $m[u, r] \gets \mathrm{Median}(\bx^{\ddagger}) - \mathrm{Median}(\bx^\star)$
            \State \com{Determine marginal gain}
            \If{$m[u, r] \ge \hat m$}
                \State $\hat m \gets m[u, r]$
                \State $(\hat u, \hat r) \gets (u, r)$
            \EndIf
        \EndFor
        \If{$\hat m > 0$}
            \State $\alpha_{\hat u} = \hat r$
            \com{$\hat u$ becomes a stooge
            with resistance $\hat r$}
        \EndIf
    \EndFor
    \State \Return $\alpha$
\EndFunction
\end{algorithmic}
\end{algorithm}

One bottleneck is that this requires
the computation of $n$ equilibrium
vectors. As discussed earlier,
computing the opinions at equilibrium
takes time $O(n^3)$, which makes for
a total running time of $O(n^4)$ per
iteration.
As such, we use an adaptation of lazy
evaluations, which are typically used
in the maximization of submodular
functions.
To describe this idea, let us
define the objective
$f(\alpha) = \mathrm{Median}(\bx^\star(\alpha, W, s))$.
We also denote with $\alpha^{(u=r)}$
the vector where we set
$\alpha^{(u=r)}_v = \alpha_v$ for $v \not= u$
and $\alpha^{(u=r)}_u = r$. In a single
iteration, we compute the marginal
gain from choosing a stooge as
$f(\alpha'^{(u=r)}) - f(\alpha)$ for each
pair $(u, r) \in V \times \{0, 1\}$,
where $\alpha$ is the resistance
in the current iteration.
However, we do not expect these
marginal gains to increase by much,
so we store them to be reused
in the coming iterations.
Specifically, in the next iteration,
we can go through all pairs
$(u, r) \in V \times \{0, 1\}$
in order decreasing with their
previous marginal gain.
Once we see that the maximum
marginal gain is larger than
all stored marginal gains by
a laziness factor of $\phi$,
we abort the search for the maximum
pair early.
The laziness $\phi$ is a
hyperparameter that we can choose.
In the best case, this helps
us to avoid the recomputation of
$n$ marginal gains per iteration.
However, this merely serves as
a heuristic, so the running time
is still $O(k n^4)$ in the worst-case.

\section{Experimental Evaluation} 
\label{sec:exp}

\subsection{Experimental Setup}

We evaluate our methods on synthetic
and real-world instances.
We wrote our code in Python~3 and ran our methods
on a 2.9 GHz Intel Xeon Gold 622R processor
with 384GB RAM.
Our code is publicly and anonymously
available~\cite{ourgit2024}.
We consider only the
maximization objective for
a threshold of $\theta = 0.5$.
This aligns with the common goal to influence a majority of votes.

\paragraph{Datasets.} We evaluate our methods on a variety of synthetic graphs, each with different topologies and $n=100$ nodes, such as grids and random binomial graphs $G(n,p)$~\cite{erdds1959random}. For simplicity,
we use uniform resistances of $\frac 1 2 $. For each network, we obtain instances by sampling innate opinions from three distinct distributions: normal, log-normal, and bimodal. Owing to space limitations, the majority of our detailed findings are presented in the appendix.


\begin{table}
  \caption{Statistics for real-world graphs. Median
  and Average refer to the equilibrium opinions $x^\star$
  before any optimization.
  We aim to maximize or minimize the median, depending
  on whether it is initially below or above the
  threshold $\theta=0.5$,
  respectively, as listed here.}
  \centering
  \medskip
  \label{tab:stats}
  \setlength{\tabcolsep}{4pt}
  \begin{tabular}{llrrrr}
    \toprule
    Instance & Reference & $n$ & $m$ & $\mathrm{Median}$ & $\mathrm{Average}$ \\
    \midrule

$\textsf{Baltimore}$&\cite{garimella18}&3902&32646&0.85&0.51 \\
$\textsf{Baltimore-F}$&&1441&28291&0.16&0.45 \\
$\textsf{Baltimore-R}$&&3902&4505&0.85&0.51 \\
$\textsf{Beefban}$&&1610&7943&0.09&0.49 \\
$\textsf{Beefban-F}$&&799&6026&0.11&0.46 \\
$\textsf{Beefban-R}$&&1610&1978&0.09&0.49 \\
$\textsf{Gunsense}$&&7106&114887&0.11&0.48 \\
$\textsf{Gunsense-F}$&&1821&103840&0.84&0.53 \\
$\textsf{Gunsense-R}$&&7106&11483&0.11&0.48 \\
$\textsf{Russia}$&&2134&19273&0.05&0.49 \\
$\textsf{Russia-F}$&&1189&16471&0.11&0.48 \\
$\textsf{Russia-R}$&&2134&2951&0.05&0.49 \\  \midrule 
$\textsf{Vax}$&\cite{ristache2024wiser}&3409&11508&0.18&0.2 \\
$\textsf{War}$&&3409&11508&0.49&0.48 \\ \midrule
$\textsf{Karate Club}$&\cite{girvan2002community}&34&78&0.46&0.47 \\

  \bottomrule
\end{tabular}
\end{table}

We also use publicly available real-world datasets
from  Garimella et al. \cite{garimella18} and Ristache et al. \cite{ristache2024wiser} (\textsf{Vax}, \textsf{Ukraine}).
The networks represent users posting on
a specific topic, and their innate opinion
$s_v \in [0, 1]$ is determined by the
sentiment expressed in a post. The
$\textsf{Vax}$ and $\textsf{War}$ datasets
use fine-grained innate opinions, while
the other datasets use
$s_v \in \{0, 1\}$. Each topic
results in two networks, where an edge exists
between two users there is a retweet
($\textsf{-R}$) or
follow relationship ($\textsf{-F}$) 
between them. We also create a third
network combining both relationships,
which we indicate by omitting
the postfix.
For the $\textsf{Vax}$ and $\textsf{War}$
datasets, both relationships are combined
into a single graph.
The resulting graphs are directed, even
though these relationships are undirected.
We also use the $\textsf{Karate Club}$ network
~\cite{girvan2002community}.
For simplicity, we assume resistances 
$\alpha_v = \frac 1 2$ for all nodes.
We show basic statistics of these datasets
in Table~\ref{tab:stats}.
We use constant weights $w_{uv} = 1$,
assuming for simplicity
that every node is uniformly
influenced by its neighbors.

\paragraph{Algorithms}

We use the algorithms introduced in
the main body and a few natural
baselines. We use the two continuous
methods $\textsf{Projected Huber}$
as described in
Algorithm~\ref{alg:huber-gd}
and the sigmoid threshold influence function
(Section~\ref{sec:sigmoid})
which we will refer to simply
as $\textsf{Sigmoid}$.
Recall that we choose resistances
to be uniformly $\frac 1 2$.
Since in our targeted problem,
the budget is on
$\|\alpha - \alpha'\|_0$,
while the budget for the continuous
problem is with respect to
$\|\alpha - \alpha'\|_1$, we
thus halve the budget in order
to compare results for discrete
targeted and continuous methods.
Instead of simple gradient
descent, we use the ADAM-Optimizer
with parameters $\beta_1 = 0.9$
and $\beta_2 = 0.999$.
We select the tuning constant $c$
for $\textsf{Projected Huber}$ via
a heuristic which we detail in
Section~\ref{subsec:ablation}.
We set a temperature
of $\tau=25$ for $\textsf{Sigmoid}$.
We use three baseline methods that
select $k$ stooges based on node
measures.
We run \textsf{Lazy Greedy}
with laziness $\phi=0.8$.
These are selecting
$k$ nodes uniformly at random
($\textsf{Random}$), selecting
the $k$ nodes with highest degree
($\textsf{Max-Degree}$) or
betweenness centrality
($\textsf{Centrality}$) \cite{freeman1977}.
For the maximization
objective, for each selected
stooge $u$, we set its resistance
to $\alpha_u = 1$ if $s_u > \theta = 0.5$
and and $\alpha_u = 0$, otherwise.

\subsection{Scalability}

We showcase the scalability of our network
on moderately sized synthetic networks and
larger real-world networks in
Figure~\ref{fig:scalability} (right).
Results on synthetic data
are in Figure~\ref{fig:scalability-synthetic}
in the appendix.
As expected, the baseline approaches
$\textsf{Random}$ and $\textsf{Max-Degree}$
are extremely fast, while $\textsf{Centrality}$
incurs some overhead to compute the centrality.
Indeed, this overhead almost approaches
the running time for the continuous approaches
$\textsf{Projected Huber}$ and $\textsf{Sigmoid}$
on larger datasets.
Both continuous approaches have very
similar running times across all datasets.
One key insight is that even though
$\textsf{Lazy Greedy}$ is fast on moderately
sized graphs, the running time cost from
recomputing marginal gains in each iteration
becomes significant a significant
bottleneck for larger graphs and
larger budgets $k$.
On the other hand, the running time
of continuous approaches does not
depend on the budget $k$, which is
noticeable on larger graphs. Our results are qualitatively consistent with the runtimes on synthetic datasets. While our work has made significant strides, scalability remains an intriguing and open direction for future research, where potentially sublinear-time algorithms for approximating functions of the equilibrium  may be helpful~\cite{neumann2024sublinear}.

\begin{figure*}[htbp]
    \centering
    \small
    \setlength{\tabcolsep}{0pt}
    \begin{tabular}{cccc}
        \includegraphics[width=0.26\textwidth]{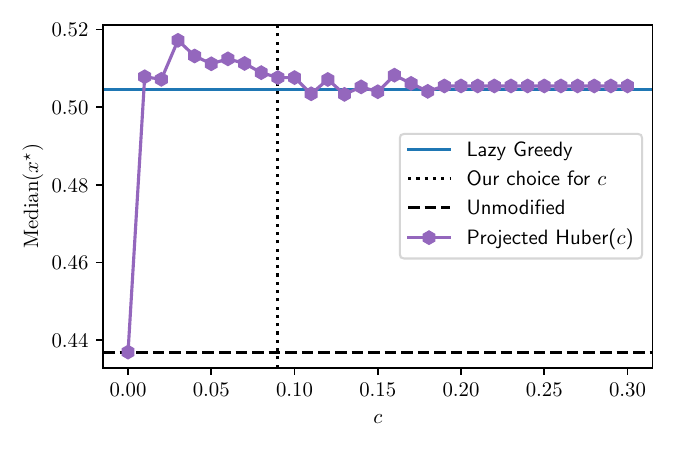} & 
        \includegraphics[width=0.26\textwidth]{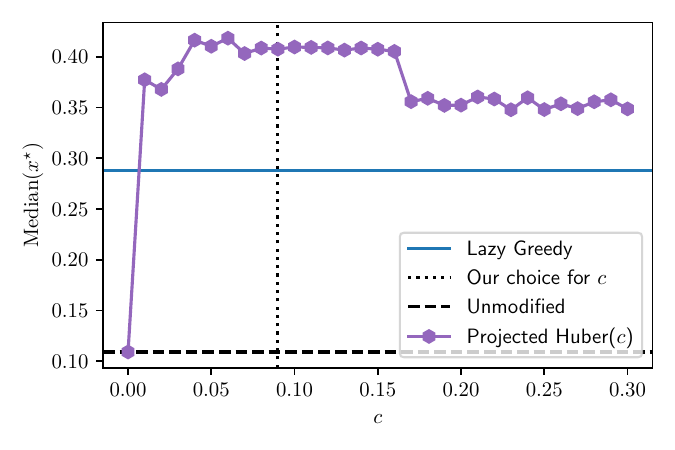}  & 
        \includegraphics[width=0.23\linewidth,trim={0 0 0 23pt},clip]{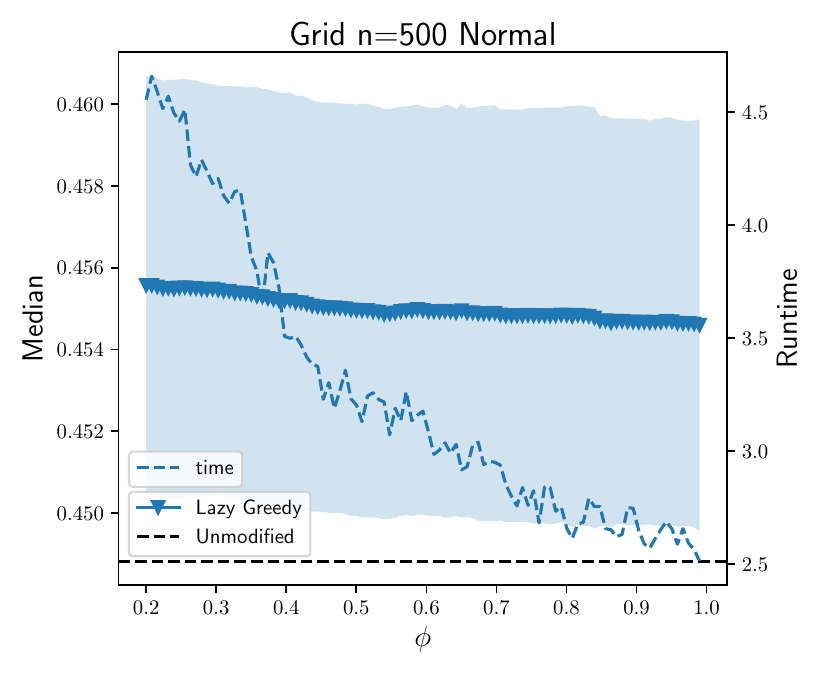} &
        \includegraphics[width=0.23\linewidth,trim={0 0 0 23pt},clip]{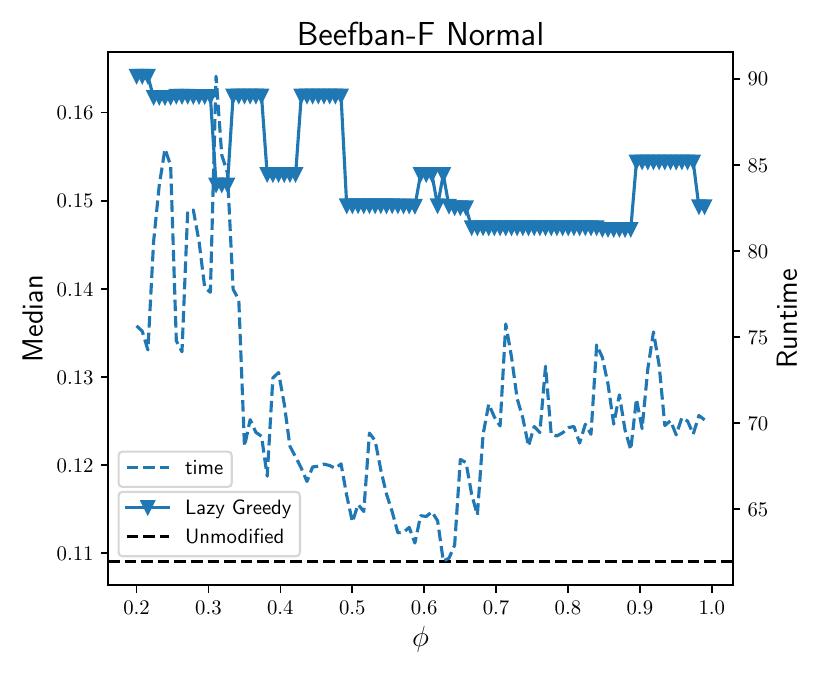} \\[-3pt]
    (a) & (b) & (c) & (d)
    \end{tabular}
    \vspace{-5pt}
    \caption{ \label{fig:ablation-c} Performance of $\textsf{Projected Huber}$ on   (a) a $10\times 10$ \textsf{Grid} and (b)  $\textsf{Beefban-F}$ with a budget of $k=50$ stooges, for various values of $c$. The plots display the instance-specific value of $c$ determined by our heuristic strategy to identify an optimal value. Median and runtime as a function of parameter $\phi$  for the $\textsf{Lazy Greedy}$ on (c) a $23 \times 23$ grid and (d) \textsf{Beefban-F}. 
    }
\end{figure*}

\subsection{Swaying the Election Results}

\begin{figure}
    \centering
    \includegraphics[width=0.5\linewidth]{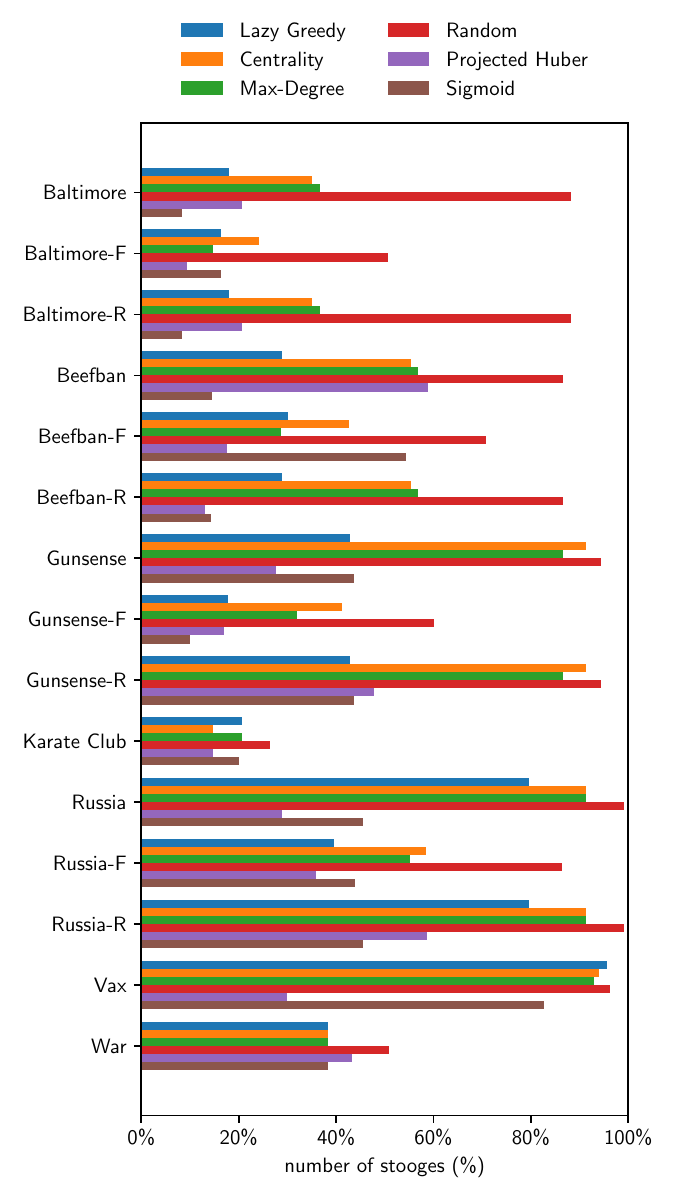}~
    \includegraphics[width=0.5\linewidth]{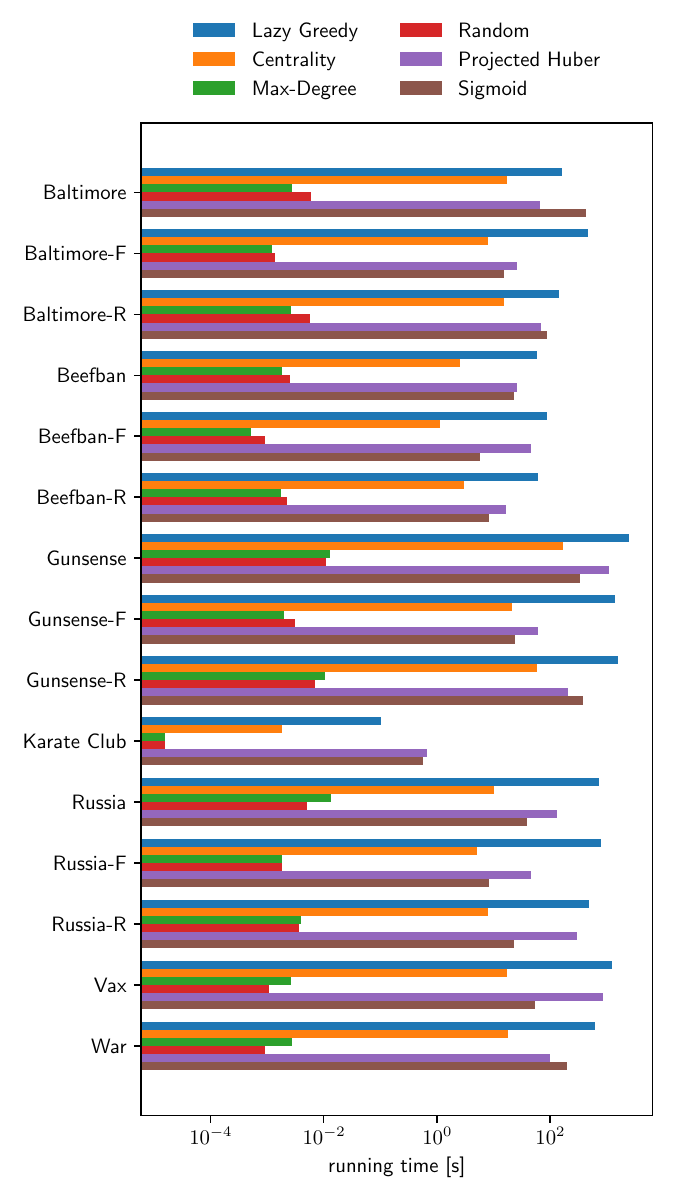}
    \figsp
    \caption{Flipping the median on real-world networks. 
    On the left,
    we show the required number of stooges to flip the median across the
    threshold of $\theta=0.5$, as a percentage of the
    number of vertices in the graph.
    On the right, we show the running times, where the budget
    is the least budget to move the median opinion above
    the threshold, corresponding to the values on the left.
    }
    \label{fig:real-world}
    \label{fig:scalability}
\end{figure}

Figure~\ref{fig:real-world} (left) shows
our results on real-world networks.
Our results show that a
small percentage of stooges suffices
to change the median drastically.
One reason for the increased
susceptibility in real-networks
over synthetic networks (cf. Figure~\ref{fig:synthetic-normal}
in the appendix) is the
high skewness: Table~\ref{tab:stats}
shows that the median is usually
far from the average opinion,
which is much closer to the 
threshold $\theta=0.5$.
We see that across most
real-world networks,
$\textsf{Projected Huber}$ is
able to flip the median
with the smallest number of stooges.  Among the targeted discrete methods, $\textsf{Lazy Greedy}$
is effective, but in some
instances requires even more
stooges than the baselines
$\textsf{Centrality}$ or
$\textsf{Max-Degree}$.
Overall, targeting a set of stooges
instead of using a continuous intervention
leads to weaker results.

\subsection{Ablation Study}
\label{subsec:ablation}

The primary parameters governing our algorithms are the Huber constant $c$ and the laziness parameter $\phi$. We discuss a heuristic for finding a good value of $c$ in the sense of yielding a near-optimal value for the equilibrium median. Additionally, we demonstrate that increasing the parameter $\phi$ generally leads to speedups, although this monotonic behavior is not consistently observed.

\spara{Tuning Huber's constant $c$}
\label{sec:find-c}


Figure~\ref{fig:ablation-c} (a) and (b) show
the performance of our continuous
method $\textsf{Projected Huber}$ for
different values of $c$.
We observe that the choice
of $c$ poses a trade-off between
the approximation of the median and 
the continuity of the objective,
which is required for successful
optimization. Our instance-specific heuristic for setting $c$ (indicated by dotted lines) identifies a value of $c$ that closely approximates the optimal value, which achieves the maximum median. Practically, it is advisable to test a range of $c$ values centered around the heuristic's suggestion, such as $0.5c, c, 2c$. 
%
%
\textsf{Projected Huber} uses a short algorithm to automatically determine a value for $c$. To find the $c$ value, we randomly perturb the graph's resistances and opinions by a small constant $\epsilon$ in either
direction, generating multiple different instances of the problem. This strategy proves robust against several choices of $\epsilon$. We then select the value of $c$ that minimizes the distance of the Huber M-estimator $\hat y$ in comparison to the true median, averaged over all instances.

\paragraph{Laziness parameter $\phi$}

Figure~\ref{fig:ablation-c} (c)
and (d) show the effect of the laziness
$\phi$.
We show the running time and resulting
median opinion.
We observe that, as expected, the
running time improves for higher
values of $\phi$ but we obtain
lower values for the median opinion.
Note that the relationship between
$\phi$ and the running time is not
necessarily monotone, as selecting
a different stooges in
Algorithm~\ref{alg:lazy-greedy}
may lead to different subsequent
decisions.

\section{Conclusion}
\label{sec:concl}
In this study, we address the problem of influencing an election under the FJ opinion dynamics model, a vital concern for developing network defense strategies. We propose two innovative approaches inspired by real-world scenarios: maximizing the median opinion or altering it to sway the election, even marginally. We demonstrate that achieving our computational objectives are NP-hard and difficult to approximate. We develop three novel algorithms leveraging both continuous and discrete optimization techniques, applied across a variety of real-world and synthetic networks. Additionally, we establish that exact solutions for flipping the median are possible on rooted directed trees, which mimic hierarchical structures. Our research opens up several intriguing questions for future exploration. Is it possible to enhance the scalability of our methods using the advancements in sublinear algorithms for opinion dynamics~\cite{neumann2024sublinear}? Can we develop efficient algorithms that are parameterized by treewidth or other graph metrics?


\clearpage
\bibliographystyle{alpha}
\bibliography{ref}

\newcommand{\etalchar}[1]{$^{#1}$}
\begin{thebibliography}{BGKMT16}

\bibitem[Abe64]{abelson1964mathematical}
Robert~P Abelson.
\newblock Mathematical models of the distribution of attitudes under
  controversy.
\newblock {\em Contributions to mathematical psychology}, 1964.

\bibitem[ACFO13]{acemouglu2013opinion}
Daron Acemo{\u{g}}lu, Giacomo Como, Fabio Fagnani, and Asuman Ozdaglar.
\newblock Opinion fluctuations and disagreement in social networks.
\newblock {\em Mathematics of Operations Research}, 38(1):1--27, 2013.

\bibitem[ACK{\etalchar{+}}22]{abebe2020opinion}
Rediet Abebe, TH~Chan, Jon Kleinberg, Zhibin Liang, David Parkes, Mauro Sozio,
  and Charalampos Tsourakakis.
\newblock Opinion dynamics with varying susceptibility to persuasion via
  non-convex local search.
\newblock {\em ACM Transactions on Knowledge Discovery from Data (TKDD)}, 16,
  2022.

\bibitem[AG17]{allcott2017social}
Hunt Allcott and Matthew Gentzkow.
\newblock Social media and fake news in the 2016 election.
\newblock {\em Journal of economic perspectives}, 31(2):211--36, 2017.

\bibitem[AO11]{acemoglu2011opinion}
Daron Acemoglu and Asuman Ozdaglar.
\newblock Opinion dynamics and learning in social networks.
\newblock {\em Dynamic Games and Applications}, 1(1):3--49, 2011.

\bibitem[Asc55]{asch1955opinions}
Solomon~E. Asch.
\newblock Opinions and social pressure.
\newblock {\em Scientific American}, 193(5):31--35, 1955.

\bibitem[AW15]{abramowitz2015all}
Alan Abramowitz and Steven Webster.
\newblock All politics is national: The rise of negative partisanship and the
  nationalization of us house and senate elections in the 21st century.
\newblock In {\em Annual Meeting of the Midwest Political Science Association
  Conference}, pages 16--19, 2015.

\bibitem[BBC14]{bbc}
BBC.
\newblock America's political divide by the numbers, 2014.
\newblock
  \href{https://www.bbc.com/news/av/magazine-27629535/america-s-political-divide-by-the-numbers}{Source
  BBC}.

\bibitem[BBC16]{bbc2}
BBC.
\newblock The reasons why politics feels so tribal in 2016, 2016.
\newblock
  \href{https://www.bbc.com/future/article/20160823-how-modern-life-is-destroying-democracy}{Source
  BBC}.

\bibitem[BBPC23]{biondi2023dynamics}
Elisabetta Biondi, Chiara Boldrini, Andrea Passarella, and Marco Conti.
\newblock Dynamics of opinion polarization.
\newblock {\em {IEEE} Trans. Syst. Man Cybern. Syst.}, 53(9):5381--5392, 2023.

\bibitem[BGKMT16]{berenbrink2016bounds}
Petra Berenbrink, George Giakkoupis, Anne-Marie Kermarrec, and Frederik
  Mallmann-Trenn.
\newblock Bounds on the voter model in dynamic networks.
\newblock {\em arXiv preprint arXiv:1603.01895}, 2016.

\bibitem[BGS20]{boxell2020cross}
Levi Boxell, Matthew Gentzkow, and Jesse~M Shapiro.
\newblock Cross-country trends in affective polarization.
\newblock Technical report, National Bureau of Economic Research, 2020.

\bibitem[BHY01]{brown2001smoothed}
BM~Brown, Peter Hall, and GA~Young.
\newblock The smoothed median and the bootstrap.
\newblock {\em Biometrika}, 88(2):519--534, 2001.

\bibitem[BKO15]{bindel2015bad}
David Bindel, Jon Kleinberg, and Sigal Oren.
\newblock How bad is forming your own opinion?
\newblock {\em Games and Economic Behavior}, 92:248--265, 2015.

\bibitem[CFL09]{castellano2009statistical}
Claudio Castellano, Santo Fortunato, and Vittorio Loreto.
\newblock Statistical physics of social dynamics.
\newblock {\em Reviews of modern physics}, 81(2):591, 2009.

\bibitem[Cia01]{cialdini2001science}
Robert~B Cialdini.
\newblock The science of persuasion.
\newblock {\em Scientific American}, 284(2):76--81, 2001.

\bibitem[CL21]{chan2021hardness}
T.{-}H.~Hubert Chan and Chui~Shan Lee.
\newblock On the hardness of opinion dynamics optimization with
  l\({}_{\mbox{1}}\)-budget on varying susceptibility to persuasion.
\newblock In {\em {COCOON}}, volume 13025 of {\em Lecture Notes in Computer
  Science}, pages 515--527. Springer, 2021.

\bibitem[CR22]{chen22}
Mayee~F. Chen and Mikl{\'{o}}s~Z. R{\'{a}}cz.
\newblock An adversarial model of network disruption: Maximizing disagreement
  and polarization in social networks.
\newblock {\em {IEEE} Trans. Netw. Sci. Eng.}, 9(2):728--739, 2022.

\bibitem[DeG74]{degroot1974reaching}
Morris~H DeGroot.
\newblock Reaching a consensus.
\newblock {\em Journal of the American Statistical Association},
  69(345):118--121, 1974.

\bibitem[Dep18]{usjustice}
U.S.~Justice Department.
\newblock Internet research agency indictment - {D}epartment of {J}ustice,
  2018.
\newblock \href{https://www.justice.gov/file/1035477/download}{URL}.

\bibitem[ER59]{erdds1959random}
P~ERDdS and A~R\&wi.
\newblock On random graphs i.
\newblock {\em Publ. math. debrecen}, 6(290-297):18, 1959.

\bibitem[FB17]{friedkin2017truth}
Noah~E Friedkin and Francesco Bullo.
\newblock How truth wins in opinion dynamics along issue sequences.
\newblock {\em Proceedings of the National Academy of Sciences},
  114(43):11380--11385, 2017.

\bibitem[FJ56]{french1956formal}
John~RP French~Jr.
\newblock A formal theory of social power.
\newblock {\em Psychological review}, 63(3):181, 1956.

\bibitem[FJ90]{friedkin1990social}
Noah~E Friedkin and Eugene~C Johnsen.
\newblock Social influence and opinions.
\newblock {\em Journal of mathematical sociology}, 15(3-4):193--206, 1990.

\bibitem[FJ97]{friedkin1997social}
Noah~E Friedkin and Eugene~C Johnsen.
\newblock Social positions in influence networks.
\newblock {\em Social networks}, 19(3):209--222, 1997.

\bibitem[FJ11]{friedkin2011social}
Noah~E Friedkin and Eugene~C Johnsen.
\newblock {\em Social influence network theory: A sociological examination of
  small group dynamics}, volume~33.
\newblock Cambridge University Press, 2011.

\bibitem[FJB16]{friedkin2016theory}
Noah~E Friedkin, Peng Jia, and Francesco Bullo.
\newblock A theory of the evolution of social power: Natural trajectories of
  interpersonal influence systems along issue sequences.
\newblock {\em Sociological Science}, 3:444--472, 2016.

\bibitem[Fre77]{freeman1977}
LC~Freeman.
\newblock A set of measures of centrality based on betweenness.
\newblock {\em Sociometry}, 1977.

\bibitem[Fri15]{friedkin2015problem}
Noah~E Friedkin.
\newblock The problem of social control and coordination of complex systems in
  sociology: A look at the community cleavage problem.
\newblock {\em IEEE Control Systems Magazine}, 35(3):40--51, 2015.

\bibitem[GKT20]{gaitonde2020adversarial}
Jason Gaitonde, Jon Kleinberg, and Eva Tardos.
\newblock Adversarial perturbations of opinion dynamics in networks.
\newblock In {\em Proceedings of the 21st ACM Conference on Economics and
  Computation}, pages 471--472, 2020.

\bibitem[GMGM18]{garimella18}
Kiran Garimella, Gianmarco De~Francisci Morales, Aristides Gionis, and Michael
  Mathioudakis.
\newblock Quantifying controversy on social media.
\newblock {\em {ACM} Trans. Soc. Comput.}, 1(1):3:1--3:27, 2018.

\bibitem[GN02]{girvan2002community}
Michelle Girvan and Mark~EJ Newman.
\newblock Community structure in social and biological networks.
\newblock {\em Proceedings of the national academy of sciences},
  99(12):7821--7826, 2002.

\bibitem[GR20]{grabisch2020survey}
Michel Grabisch and Agnieszka Rusinowska.
\newblock A survey on nonstrategic models of opinion dynamics.
\newblock {\em Games}, 11(4):65, 2020.

\bibitem[GS13]{ghaderi2013opinion}
Javad Ghaderi and R~Srikant.
\newblock Opinion dynamics in social networks: A local interaction game with
  stubborn agents.
\newblock In {\em American Control Conference (ACC), 2013}, pages 1982--1987.
  IEEE, 2013.

\bibitem[GS14]{ghaderi2014opinion}
Javad Ghaderi and Rayadurgam Srikant.
\newblock Opinion dynamics in social networks with stubborn agents: Equilibrium
  and convergence rate.
\newblock {\em Automatica}, 50(12):3209--3215, 2014.

\bibitem[GTT13]{gionis2013opinion}
Aristides Gionis, Evimaria Terzi, and Panayiotis Tsaparas.
\newblock Opinion maximization in social networks.
\newblock In {\em Proceedings of the 2013 SIAM International Conference on Data
  Mining}, pages 387--395. SIAM, 2013.

\bibitem[Hal91]{halpern1991relationship}
Joseph~Y Halpern.
\newblock The relationship between knowledge, belief, and certainty.
\newblock {\em Annals of mathematics and artificial intelligence}, 4:301--322,
  1991.

\bibitem[HHR11]{hampel2011smoothing}
Frank Hampel, Christian Hennig, and Elvezio Ronchetti.
\newblock A smoothing principle for the huber and other location m-estimators.
\newblock {\em Computational Statistics \& Data Analysis}, 55(1):324--337,
  2011.

\bibitem[HK{\etalchar{+}}02]{hegselmann2002opinion}
Rainer Hegselmann, Ulrich Krause, et~al.
\newblock Opinion dynamics and bounded confidence models, analysis, and
  simulation.
\newblock {\em Journal of Artificial Societies and Social Simulation}, 5(3),
  2002.

\bibitem[HRS19]{haidt2019dark}
Jonathan Haidt and Tobias Rose-Stockwell.
\newblock The dark psychology of social networks.
\newblock {\em The Atlantic}, pages 6--60, 2019.

\bibitem[Hun84]{hunter1984mathematical}
John~E Hunter.
\newblock {\em Mathematical Models of Attitude Change. Volume 1, Change in
  Single Attitudes and Cognitive Structure}.
\newblock 1984.

\bibitem[KKT03]{kempe2003maximizing}
David Kempe, Jon Kleinberg, and {\'E}va Tardos.
\newblock Maximizing the spread of influence through a social network.
\newblock In {\em Proceedings of the ninth ACM SIGKDD international conference
  on Knowledge discovery and data mining}, pages 137--146. ACM, 2003.

\bibitem[Kle20]{klein2020we}
Ezra Klein.
\newblock {\em Why we're polarized}.
\newblock Simon and Schuster, 2020.

\bibitem[Lor07]{lorenz2007continuous}
Jan Lorenz.
\newblock Continuous opinion dynamics under bounded confidence: A survey.
\newblock {\em International Journal of Modern Physics C}, 18(12):1819--1838,
  2007.

\bibitem[LW19]{rolling}
Darren Linvill and Patrick Warren.
\newblock That uplifting tweet you just shared? a russian troll sent it, 2019.
\newblock
  \href{https://www.rollingstone.com/politics/politics-features/russia-troll-2020-election-interference-twitter-916482/}{URL}.

\bibitem[LZ19]{ledwich2019algorithmic}
Mark Ledwich and Anna Zaitsev.
\newblock Algorithmic extremism: Examining youtube's rabbit hole of
  radicalization.
\newblock {\em arXiv preprint arXiv:1912.11211}, 2019.

\bibitem[McN20]{mcnamee2020zucked}
Roger McNamee.
\newblock {\em Zucked: Waking up to the Facebook catastrophe}.
\newblock Penguin, 2020.

\bibitem[MMT18]{musco18}
Cameron Musco, Christopher Musco, and Charalampos~E. Tsourakakis.
\newblock Minimizing polarization and disagreement in social networks.
\newblock In {\em {WWW}}, pages 369--378. {ACM}, 2018.

\bibitem[NDP24]{neumann2024sublinear}
Stefan Neumann, Yinhao Dong, and Pan Peng.
\newblock Sublinear-time opinion estimation in the friedkin--johnsen model.
\newblock In {\em Proceedings of the ACM on Web Conference 2024}, pages
  2563--2571, 2024.

\bibitem[Obs20]{astroturf}
Stanford~Internet Observatory.
\newblock Reply-guys go hunting: An investigation into a {U}.{S}. astroturfing
  operation on facebook, twitter, and instagram.
\newblock 2020.
\newblock
  \href{https://stacks.stanford.edu/file/druid:vh222ch4142/facebook-US-202009.pdf}{URL}.

\bibitem[our24]{ourgit2024}
Anonymous repository, 2024.
\newblock Available at \url{https://anonymous.4open.science/r/Elections-FB8B}.

\bibitem[PT17]{proskurnikov2017tutorial}
Anton~V Proskurnikov and Roberto Tempo.
\newblock A tutorial on modeling and analysis of dynamic social networks. part
  i.
\newblock {\em Annual Reviews in Control}, 43:65--79, 2017.

\bibitem[PT18]{proskurnikov2018tutorial}
Anton~V Proskurnikov and Roberto Tempo.
\newblock A tutorial on modeling and analysis of dynamic social networks. part
  ii.
\newblock {\em Annual Reviews in Control}, 45:166--190, 2018.

\bibitem[RD02]{richardson2002mining}
Matthew Richardson and Pedro Domingos.
\newblock Mining knowledge-sharing sites for viral marketing.
\newblock In {\em Proceedings of the eighth ACM SIGKDD international conference
  on Knowledge discovery and data mining}, pages 61--70, 2002.

\bibitem[Res14]{pew2}
Pew Research.
\newblock Political polarization in the american public, 2014.
\newblock
  \href{https://www.people-press.org/2014/06/12/political-polarization-in-the-american-public/}{Pew
  Research link}.

\bibitem[Res16]{pew1}
Pew Research.
\newblock Partisanship and political animosity in 2016, 2016.
\newblock
  \href{https://www.people-press.org/2016/06/22/partisanship-and-political-animosity-in-2016/}{Pew
  Research link}.

\bibitem[RR23]{DBLP:journals/corr/abs-2206-08996}
Mikl{\'{o}}s~Z. R{\'{a}}cz and Daniel~E. Rigobon.
\newblock Towards consensus: Reducing polarization by perturbing social
  networks.
\newblock {\em {IEEE} Trans. Netw. Sci. Eng.}, 10(6):3450--3464, 2023.

\bibitem[RST24]{ristache2024wiser}
Dragos Ristache, Fabian Spaeh, and Charalampos~E Tsourakakis.
\newblock Wiser than the wisest of crowds: The asch effect revisited under
  friedkin-johnsen opinion dynamics.
\newblock {\em arXiv e-prints}, pages arXiv--2406, 2024.

\bibitem[SG19]{smith2019mapping}
Naomi Smith and Tim Graham.
\newblock Mapping the anti-vaccination movement on facebook.
\newblock {\em Information, Communication \& Society}, 22(9):1310--1327, 2019.

\bibitem[SZ23]{sun2023opinion}
Haoxin Sun and Zhongzhi Zhang.
\newblock Opinion optimization in directed social networks.
\newblock In {\em {AAAI}}, pages 4623--4632. {AAAI} Press, 2023.

\bibitem[TBA86]{tsitsiklis1986distributed}
John Tsitsiklis, Dimitri Bertsekas, and Michael Athans.
\newblock Distributed asynchronous deterministic and stochastic gradient
  optimization algorithms.
\newblock {\em IEEE transactions on automatic control}, 31(9):803--812, 1986.

\bibitem[Tim16]{nytimes}
New~York Times.
\newblock How large is the divide between red and blue america?, 2016.
\newblock
  \href{https://www.nytimes.com/interactive/2016/11/04/us/politics/growing-divide-between-red-and-blue-america.html
  }{Source NY Times}.

\bibitem[Tim18]{nix}
New~York Times.
\newblock How {T}rump consultants exploited the {F}acebook data of millions,
  2018.
\newblock
  \href{https://www.nytimes.com/2018/03/17/us/politics/cambridge-analytica-trump-campaign.html}{URL}.

\bibitem[TTZ{\etalchar{+}}21]{tang2021susceptible}
Wenyi Tang, Ling Tian, Xu~Zheng, Guangchun Luo, and Zaobo He.
\newblock Susceptible user search for defending opinion manipulation.
\newblock {\em Future Generation Computer Systems}, 115:531--541, 2021.

\bibitem[WS11a]{williamson2011design}
David~P Williamson and David~B Shmoys.
\newblock {\em The design of approximation algorithms}.
\newblock Cambridge university press, 2011.

\bibitem[WS11b]{williamsonbook}
David~P Williamson and David~B Shmoys.
\newblock {\em The design of approximation algorithms}.
\newblock Cambridge university press, 2011.

\bibitem[YOA{\etalchar{+}}13]{yildiz2013binary}
Ercan Yildiz, Asuman Ozdaglar, Daron Acemoglu, Amin Saberi, and Anna Scaglione.
\newblock Binary opinion dynamics with stubborn agents.
\newblock {\em ACM Transactions on Economics and Computation (TEAC)},
  1(4):1--30, 2013.

\bibitem[ZBZ21]{zhu2021minimizing}
Liwang Zhu, Qi~Bao, and Zhongzhi Zhang.
\newblock Minimizing polarization and disagreement in social networks via link
  recommendation.
\newblock In {\em NeurIPS}, pages 2072--2084, 2021.

\bibitem[ZZ22]{zhu2022nearly}
Liwang Zhu and Zhongzhi Zhang.
\newblock A nearly-linear time algorithm for minimizing risk of conflict in
  social networks.
\newblock In {\em KDD}, pages 2648--2656, 2022.

\end{thebibliography}

\clearpage
\appendix

\label{sec:appendix}

\section{Additional Experimental Results}

We now provide further details and results
from Section~\ref{sec:exp}, which we omitted
due to space limitations from the main body.
We begin by providing further details
of the sigmoid threshold influence function
and then provide omitted details
and results for synthetic instances.

\spara{Sigmoid threshold influence function}

\begin{algorithm}
\caption{Flipping the Median via the Sigmoid Threshold Influence Function.}
\label{alg:sigmoid}
\begin{algorithmic}[1]
\Function{SigmoidGD}{$G, \alpha_0, s, k, \eta$}
    \State $\alpha \gets \alpha_0$
    \While{not converged}
        \State Let $X = I - (I - A) M$ where $A = \mathrm{Diag}(\alpha)$
        \State Solve $x^\star = \min_x \|X x - A s \|_2$
        \com{Calculate opinions}
        \State $\alpha' \gets \alpha + \eta \nabla_\alpha f_{\mathrm{sigmoid}}(x^\star)$
        \com{Gradient update}
        \State $\alpha \gets \min \{ \|\alpha - \alpha'\|_2 :
            \|\alpha - \alpha_0\|_1 \le k \}$
        \com{Projection}
    \EndWhile
    \State \Return $\alpha$
\EndFunction
\end{algorithmic}
\end{algorithm}

In Algorithm~\ref{alg:sigmoid},
we provide the full pseudocode for the
continuous method based on the sigmoid threshold
influence function
as described in Section~\ref{sec:sigmoid}.
We can easily compute
the gradient of the sigmoid method via
the chain rule and the gradient
for $x^\star$ which we determined in
Section~\ref{sec:proposed}:
\begin{align*}
    \nabla_{\alpha} f_{\mathrm{sigmoid}}(x^\star) = \tau \mathrm{Diag}(s - W x^\star)
    (X^+)^\top \sigma
\end{align*}
where $\sigma \in \R^{n}$
is the vector
with $\sigma_u = \mathrm{sigmoid}(x_u^\star)$.
Again, we can avoid computing
the pseudoinverse and instead
solve the least squares
problem $\min_z \| X^\top z - \sigma \|_2$.

\spara{Setup of our synthetic datasets}

We use various synthetic graphs on
different topologies with $n=100$ nodes
to capture a wide range of possible
connections and edge test cases. Our deterministic instances
are the $10 \times 10$ grid graph 
$\textsf{Grid}$ and
the star graph $\textsf{Star}$ with
one center node and
$99$ leaves. We use the
$\textsf{GNP}(n, 0.05)$ model
\cite{erdds1959random}
where each pair of vertices
is independently included as an
edge with probability $0.05$ for sufficiently large $n$ that ensures connectivity.
In such a random graph, all
vertices are essentially identical,
which makes every node equally
(in)-effective as a stooge.
%
We denote with $\textsf{Tree}$
the instance where we sample
uniformly from the set of all
trees on $n$ nodes. Finally,
we use a preferential attachment
model $\textsf{BA}$ where we add
one node after the other to the
network, whereas we connect
each node to $5$ randomly to
$5$ other nodes with probability
proportional to their current
degree.
Finally, we use a stochastic
block model $\textsf{Communities}$
consisting of a big community
with $50$ nodes and $5$ communities
of $10$ nodes each. The probability
for inter-community edges is $0.3$
and for intra-community edges $0.5$.
We sample innate opinions from three
different distributions with different
concentration properties. We use
a normal distribution
($\textsf{Normal}$)
and 
a log-normal distribution for
($\textsf{LogNormal}$)
to simulate both concentrated
and anti-concentrated opinions.
In both cases, we set parameters
such that the mean is $\mu=0.45$
and the standard deviation
$\sigma=0.1$, and we then truncate
them to the range $[0, 1]$.
We also use a bimodal distribution
where we toss a fair coin,
and assign the innate opinion $0.35$
if it comes up heads and $0.55$
if it comes up tails.
This choice ensures that mean
and standard deviation
across all distributions is
identical.

\begin{figure}
    \centering
    \includegraphics[width=0.5\linewidth]{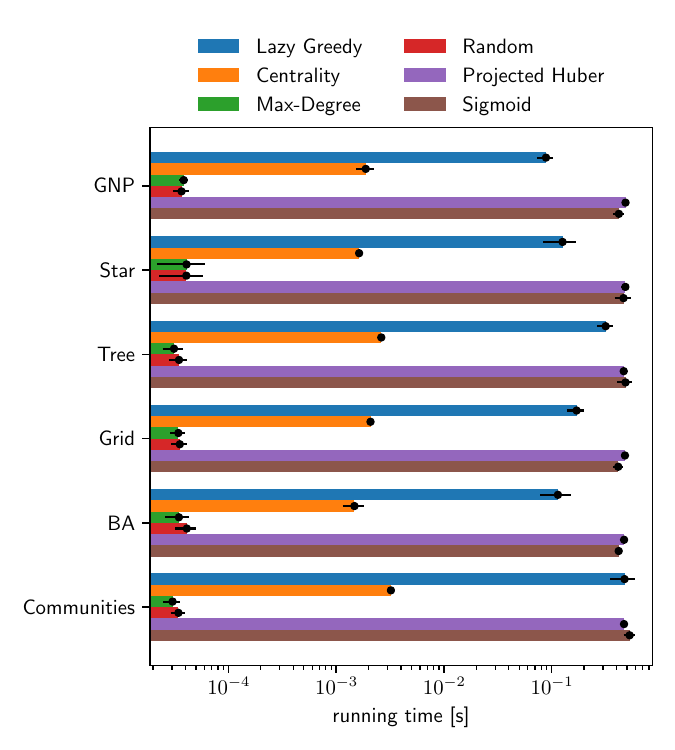}
    \caption{Scalability on synthetic
    networks when innate opinions follow
    a $\textsf{Normal}$ distribution.}
    \label{fig:scalability-synthetic}
\end{figure}

\begin{figure}
    \centering
    \includegraphics[width=0.5\linewidth]{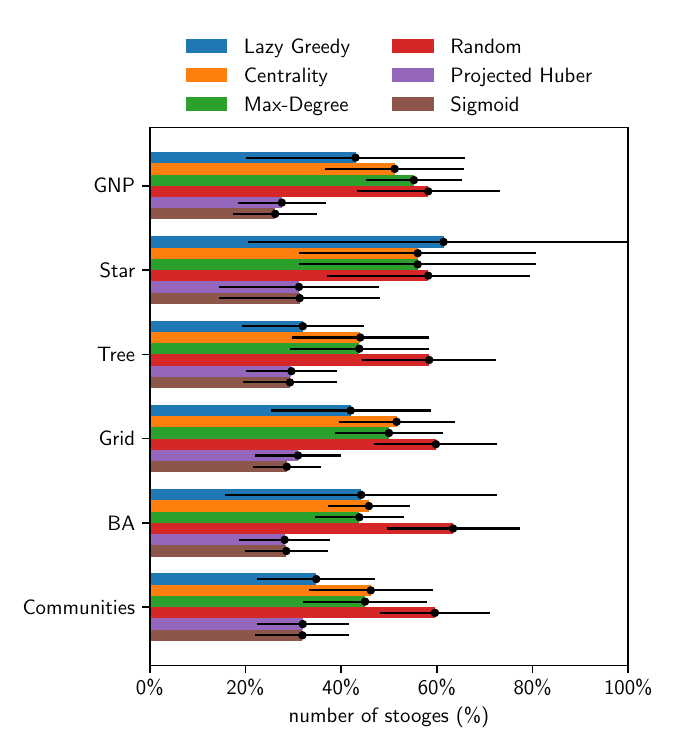}
    \figsp
    \caption{
    \label{fig:synthetic-normal}
    Median maximization on synthetic graphs
    with $\textsf{Normal}$ innate opinions.
    We show the number
    of stooges required to move the median
    over the threshold $\theta = 0.5$, as a percentage of the number of nodes in the graph. We report average and standard deviation over 10 runs.}
\end{figure}

\begin{figure}
    \centering
    \includegraphics[width=0.5\linewidth]{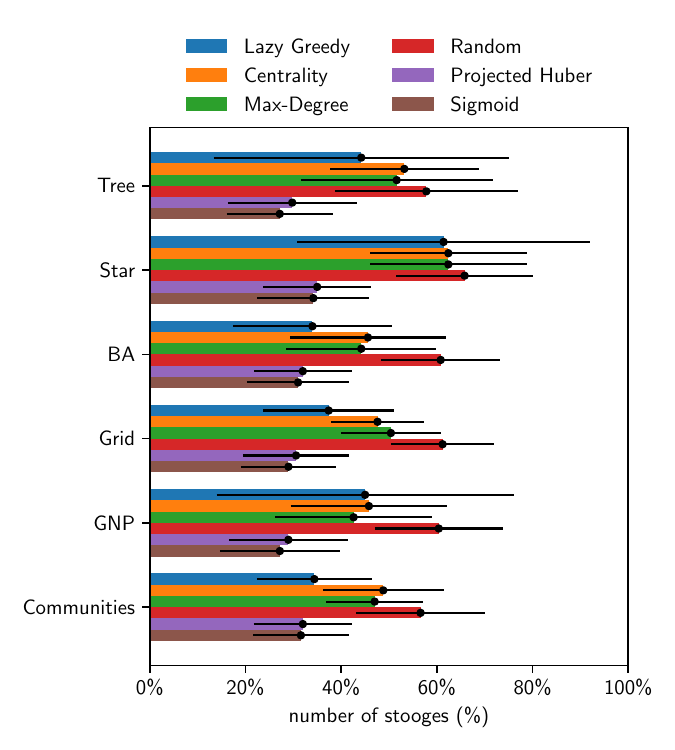}
    \figsp
    \caption{
    \label{fig:synthetic-log-normal}
    Median maximization on synthetic graphs
    with $\textsf{LogNormal}$ innate opinions.
    The legend is as in Figure~\ref{fig:synthetic-normal}}
\end{figure}

\begin{figure}
    \centering
    \includegraphics[width=0.5\linewidth]{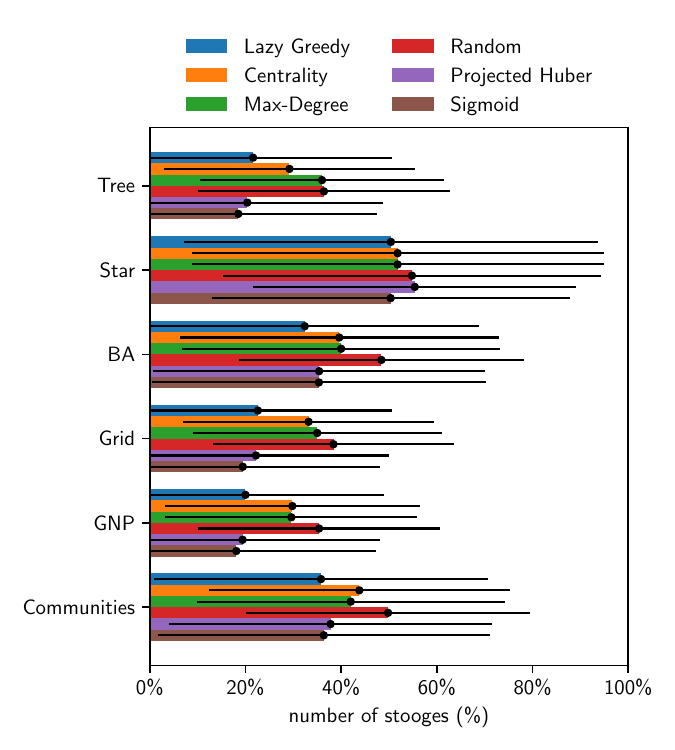}
    \figsp
    \caption{
    \label{fig:synthetic-bimodal}
    Median maximization on synthetic graphs
    with $\textsf{Bimodal}$ innate opinions.
    The legend is as in Figure~\ref{fig:synthetic-normal}}
\end{figure}

\begin{figure}
    \centering
    \includegraphics[width=0.5\linewidth]{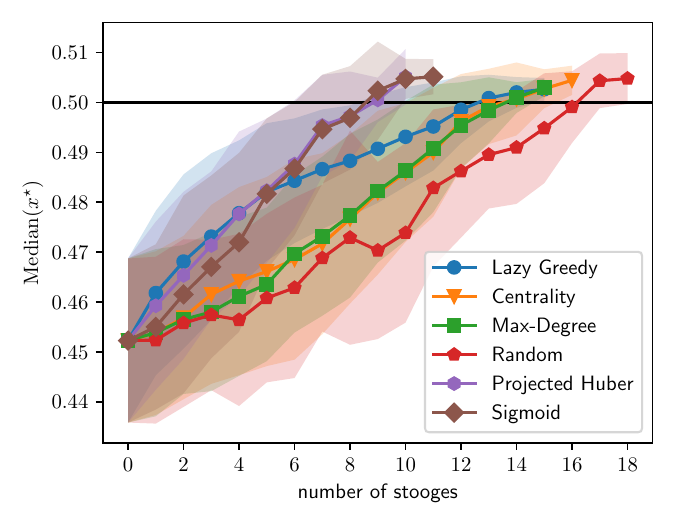}
    \figsp
    \caption{Flipping the median opinion
    on a $\textsf{Grid}$. We show the
    median opinion after optimization
    for an increasing number of stooges,
    until the threshold of $\theta=0.5$
    is reached.}
    \label{fig:max-grid}
\end{figure}

\spara{Results on synthetic datasets}

Figure~\ref{fig:synthetic-normal} 
shows how many
stooges are required to flip the median
above the threshold $\theta=0.5$,
for our methods and baselines.
Figure~\ref{fig:synthetic-normal} shows only
results for synthetic networks where innate
opinions are distributed according to
$\textsf{Normal}$. Results for
$\textsf{LogNormal}$ and $\textsf{Bimodal}$
innate opinions are in
Figures~\ref{fig:synthetic-log-normal}
and ~\ref{fig:synthetic-bimodal}, respectively.
We can observe that the continuous methods
$\textsf{Projected Huber}$ and $\textsf{Sigmoid}$
are the most efficient and require the
least budget.
Interestingly, continuous
methods perform relatively constant across
all topologies, while discrete methods
clearly depend on the graph topology.
For instance, targeted attacks via
discrete methods on a centralized graphs
$\mathsf{Star}$ or the $\mathsf{Communities}$
require the least stooges.
Interestingly, $\textsf{Tree}$ networks
require a high amount of targeted
stooges. Opinions for each
node are sampled independently, and
the few central and thus highly
influential nodes in a tree are not
guaranteed to have an innate
opinion $> 0.5$.
Figure~\ref{fig:max-grid}
shows the progress of the median
opinion for an increasing number
of stooges. We observe that
$\textsf{Lazy Greedy}$ and
$\textsf{Projected Huber}$
are both able to increase
the median a lot for a low
number of stooges. However,
the two continuous method
$\textsf{Lazy Greedy}$
and $\textsf{Sigmoid}$
are able to eventually flip the median
with the least stooges.
It is expected that
$\textsf{Sigmoid}$ is not competitive
until the stooges are able to flip
the median, as it is solely designed
for Problem~\ref{prob:elections}.

\begin{figure}
    \centering
    \includegraphics[width=0.5\linewidth]{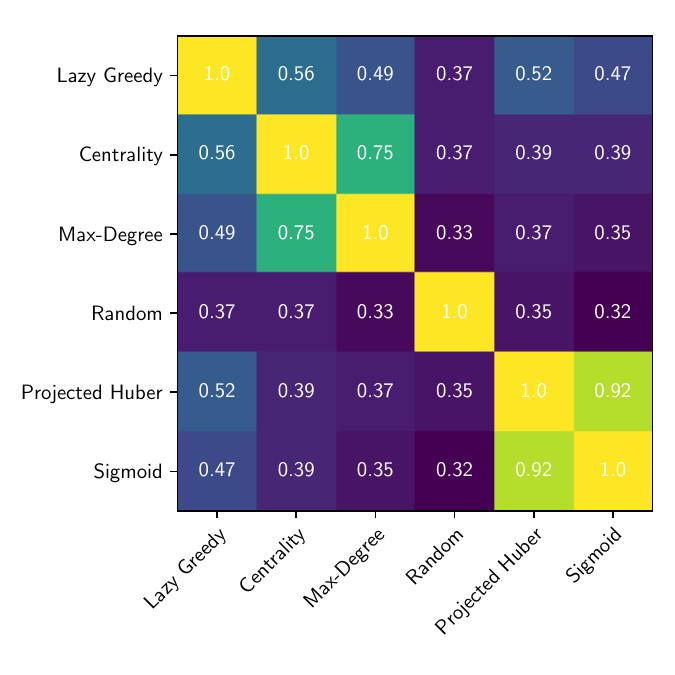}~
    \includegraphics[width=0.5\linewidth]{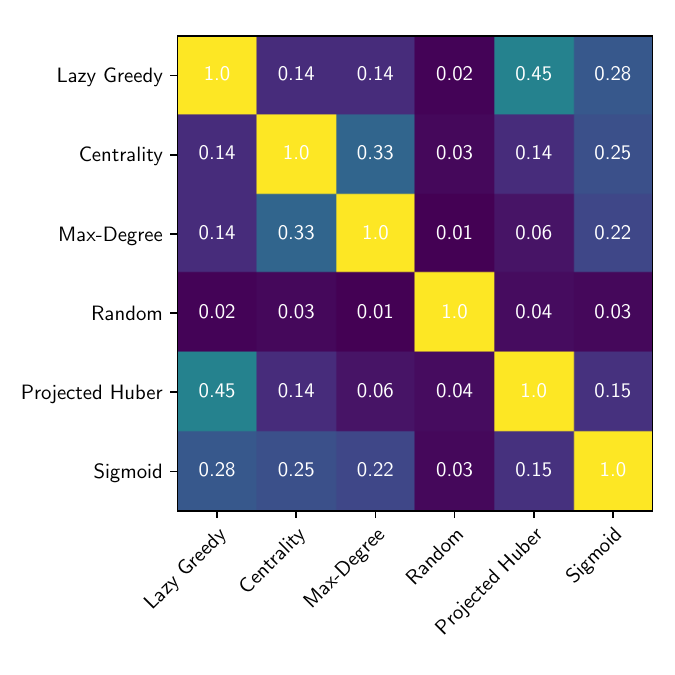}
    \figsp
    \caption{Pairwise Jaccard similarity on the set of
    selected $k=50$ stooges that are selected by each algorithm, on
    a $10 \times 10$ \textsf{Grid} for a single
    instance with \textsf{Normal} innate opinions (top)
    and \textsf{Beefban-F} (bottom).}
    \label{fig:jaccard}
\end{figure}

We also investigate
similarities between our
approaches and baselines in
Figure~\ref{fig:jaccard}.
Here, we report the
Jaccard similarity between
sets of stooges
$U_1, U_2 \subseteq V$ which
is defined as the ratio
$J(U_1, U_2) = |U_1 \cap U_2| / |U_1 \cup U_2|$.
For continuous approaches,
we select the $k$ vertices
whose resistance deviates
the most from the initial
resistance of $0.5$.
We observe that on a synthetic
$10\times 10$ \textsf{Grid}, \textsf{Projected Huber}
and \textsf{Sigmoid} perform
similarly. This is, however,
not the case for the real-world
graph \textsf{Beefban-F} where
\textsf{Projected Huber}
is more similar to \textsf{Lazy Greedy}
in its choice of stooges.

\section{Hierarchy Graphs} 
\label{sec:treedp}

\begin{figure}
    \centering
    \includegraphics[width=0.5\linewidth]{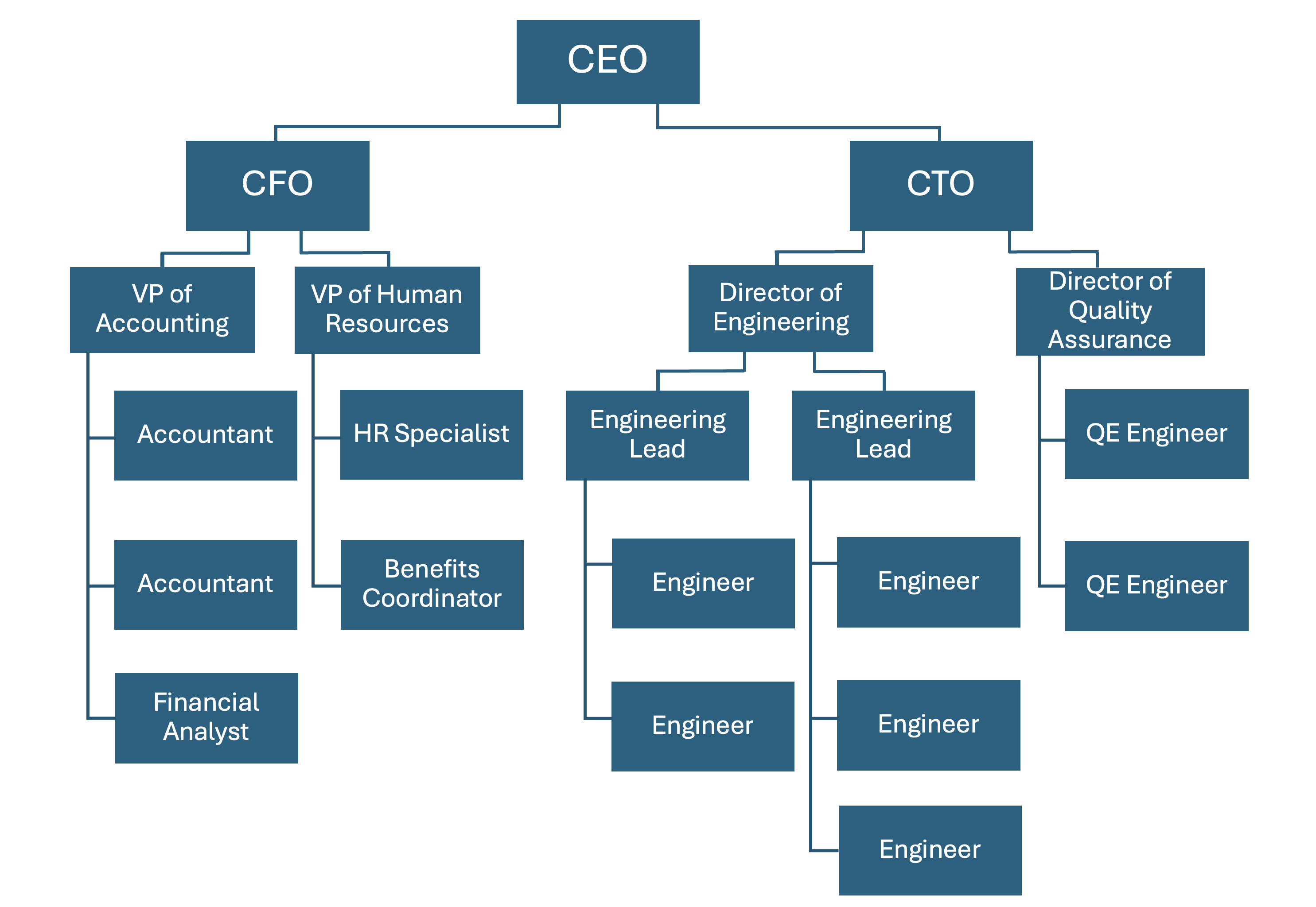}
    \figsp
    \caption{An organizational chart, which is an example of a real world hierarchy graph, where the information only flows from each person to their superior.}
    \label{fig:org-chart}
\end{figure}

In this section, we present a polynomial time algorithm that finds the optimal solution for hierarchy graphs. A hierarchy graph is a rooted directed tree where all the edges are pointing away from the root. An example is shown Figure~\ref{fig:org-chart}. Such graphs model organizational hierarchies. 
In this type of tree, information flows from bottom up, which is a natural representation of multiple real life scenarios, such as people on the lower levels in the hierarchy reporting what their opinion to their immediate superior. 

While we presented the FJ model for undirected graphs, it naturally adapts to directed graphs as follows:  
\begin{align*}
\textstyle x_u(t+1) = \alpha_u s_u + \frac{1-\alpha_u}{\deg(u)} \sum_{v \in N_{\mathrm{out}}(u)}  w_{uv} x_v(t) \hfill \text{~for all~~}u \in V.
\end{align*}
\subsection{General approach}

Formally, a hierarchy graph is a rooted tree $T$ where $w_{uv} > 0$ only if $v$ is the direct child of $u$.

As a result, each node is only influenced by the nodes in its subtree, so we can compute its equilibrium opinion by recursively computing the equilibrium opinions of its children, which are independent from each other. We use a dynamic programming approach to ensure we are getting an optimal solution. 
%
The other reason why we can achieve polynomial time is because the continuous values that represent the opinion of each node become discrete at the end by being either a \textsf{yes} vote or a \textsf{no} vote,  depending on whether they are above $0.50$ or not respectively.  
%

\subsection{Dynamic programming algorithm}

Now we explain how to use dynamic programming to find the optimal choice of stooges to bring the opinion of more than half the nodes above $0.5$.

Let $\text{dp}_{u,\text{numVotes},k}  $ denote the maximum opinion for node $u$ when the subtree is rooted in $u$ so that $k$ nodes are stooges and we have $\text{numVotes}$ nodes that have opinion $>0.5$.
The optimal number of stooges will be the smallest $k$ such that there exists a $j > \frac{n}{2}$ so that $\text{dp}_{u, j, k}$ is a reachable state, where $u$ is the root. If this state is reachable it means we have a way of assigning $k$ stooges so we can get $j$ nodes with opinion $>0.5$.

\begin{align*}
    D(u) &= \textrm{descendants of $u$ in $G$} \\
    X_{u, k} &= \{ x^\star(\alpha', W, s) :
        S \subseteq D(u) \textrm{ such that } |S| = k
        \textrm{ and } \alpha'_w = \alpha_w \textrm{ for all } w \notin S \} \\
    X_{u, j, k} &= \{ x^\star \in X_{u, k} : |\{ w \in D(u)
        : x^\star_w > {\textstyle \frac 1 2} \}| = j \} \\
    \mathrm{dp}_{u, j, k} &= \left\{ \begin{array}{ll}
         \max\{ x^\star_u : x^\star \in X_{u, j, k} \}
         & \textrm{if } X_{u, j, k} \not= \emptyset \\
         \bot & \textrm{otherwise}
    \end{array} \right.
\end{align*}

The base cases are the leaves. For a leaf $\ell$ it doesn't matter if a leaf is a stooge or not since it will have the same opinion. So this corresponds to either $\text{dp}_{\ell, 1, 0} = s_\ell$ if $s_\ell > 0.5$ or $\text{dp}_{\ell, 0, 0} = s_\ell$ otherwise.

\begin{align*}
    \textrm{for a leaf $u$ and $j, k \in \{0, 1\}$ we set} \qquad
    \mathrm{dp}_{u, j, k} = \left\{ \begin{array}{ll}
         s_u & \textrm{if } j > 0 \textrm{ and } s_u > \frac 1 2 \\
        \bot & \textrm{otherwise}
    \end{array} \right.
\end{align*}

For nodes with  children, we must optimally allocate  the   $k$ available stooges among them. We do so by creating an auxiliary dynamic program: let $\text{cdp}_{i,j,k} = $ be the largest sum of opinions across the first $i$ children with a total of $j$ nodes having opinion $> 0.5$ and using a total of $k$ stooges. This is very similar to a 2D version of the knapsack problem \cite{williamsonbook}, since when we add the $(i+1)^{\text{th}}$ child and have to combine a $\text{cdp}_{i,j,k}$ with $\text{dp}_{c,j',k'}$ we create the state $\text{cdp}_{i+1,j+j',k+k'} = \text{cdp}_{i,j,k} + w_{i,c} \cdot \text{dp}_{c, j', k'}$. After these auxiliary states are calculated, we can update the original dynamic program, and check whether the current node's opinion is above $0.5$, in which case we add $1$ to the number of votes. 
The complete pseudocode is given in Algorithm~\ref{alg:algo1}.

For each node, we create an auxiliary dynamic program $\mathrm{cdp}_{i,j,k}$. Let
$D(u, i)$ be the descendants of the first $i$ children of $u$.
\begin{align*}
    Y_k &= \{ x^\star(\alpha', W, s) : S \subseteq D(u, i) \textrm{ such that } |S| = k
        \textrm{ and } \alpha'_w = \alpha_w \textrm{ for all } w \notin S \} \\
    Y_{j, k} &= \{ x^\star \in Y_k : |\{ w \in D(u, i) :
        x_w^\star > {\textstyle \frac 1 2} \}| = j \} \\
    \mathrm{cdp}_{i, j, k} &= \left\{ \begin{array}{ll}
        \max \{ \sum_{v \in D(u, i) \cap N(u)} w_{uv} x^\star_v : x^\star \in Y_{j, k} \} & \textrm{if } Y_{j, k} \not= \emptyset \\
        \bot & \textrm{otherwise}
    \end{array} \right.
\end{align*}

The overall complexity is the same as the auxiliary dynamic program, which is $O(n^4)$ per child node, for a total of $O(n^5)$.

\begin{algorithm}
\caption{Dynamic Program for Selecting Stooges in Hierarchy Graphs}\label{alg:algo1}
\begin{algorithmic}[1]
\Function{ComputeDP}{$\mathrm{dp}, u, G$}
    \If{$u$ is a leaf}
        \If{$s_x > 0.5$}
            \State $\mathrm{dp}_{u,1,0} = s_u$
        \Else
            \State $\mathrm{dp}_{u,0,0} = s_u$
        \EndIf
    \EndIf
    \For{$v \in N_{\mathrm{out}}(u)$}
        \State $\textsc{ComputeDP}(v, G)$
    \EndFor
    \State $\textrm{cdp}_{i,j,k} \leftarrow -\infty$
    \State $\textrm{cdp}_{0,0,0} = 0$
    \For{$i = 1, \dots, \deg_{\mathrm{out}}(u)$}   
        \com{Iterate through children}
        \State Let $c$ be the $i$-th child of $u$
        \For{$\mathrm{pn} = 1, \dots, n$}
            \com{Fix number of nodes with opinion $>0.5$}
            \For{$\mathrm{sc} = 1, \dots, n$}
            \com{Fix number of stooges}
                \For{$\mathrm{pn}_c = 1, \dots, n$}
                \com{Fix \#descendants of $c$ with opinion $>0.5$}
                    \For{$\mathrm{sc}_c = 1, \dots, n$}
                    \com{Fix \#descendants of $c$ that are stooges}
                        \State $\textrm{cpd}_{i, \mathrm{pn}, \mathrm{sc}} \leftarrow \max(\mathrm{cpd}_{i, \mathrm{pn}, \mathrm{sc}}, \mathrm{dp}_{c, pb_c, \mathrm{sc}_c} \cdot \ w_{uc} + \textrm{cdp}_{i-1, \mathrm{pn}-\mathrm{pn}_c, \mathrm{sc}-\mathrm{sc}_c})$
                    \EndFor
                \EndFor
            \EndFor
        \EndFor
    \EndFor
    \For{$j = 1, \dots, n$}
        \For{$k \in 0, 1, \dots, n$}
        \If{\textrm{cdp}[$\deg_{\mathrm{out}}(u)][j][k]) = -\infty$}
            \State {\bf continue}
        \EndIf
        \State $\mathrm{pos} = 0$
        \com{If current node has opinion $>0.5$}
        \State $\textrm{finalOpinion}\leftarrow w_{u,u} \cdot s_u + \textrm{cdp}_{\deg_{\mathrm{out}}(u), j, k}$
        \If {$\textrm{finalOpinion} \ge 0.5$}
            \State $\mathrm{pos} = 1$
        \EndIf
        \State $\mathrm{dp}_{u,j+\mathrm{pos},k} \leftarrow \textrm{finalOpinion}$
        \com{If $u$ is not a stooge}
        \State $\mathrm{dp}_{u,j+1,k+1} \leftarrow \max(\mathrm{dp}_{u,j+\mathrm{pos},k},\textrm{$s_u$})$
        \com{If $u$ is a stooge}
        \EndFor
    \EndFor
\EndFunction
\end{algorithmic}
\end{algorithm}

\subsection{Extensions}

In this section we explain in detail how to change the implementation to accommodate extensions:

\textit{Only some of the nodes have voting power}: To make nodes take into account who is able to vote, we can set the variable $\text{positive}$ to $1$ in the last section of the algorithm only if the current node is a voting node. This will ensure that the algorithm only counts the voting nodes.

\textit{Different nodes have different integer costs to become stooges}: If we want to add a cost per node, we can make the third state of our dynamic programming signify the total cost rather than the total number of stooges. Most of the algorithm can stay the same, with the difference that instead of increasing by 1 if we set node $x$ to be a stooge, we increase it by $\text{cost}_x$, where $\text{cost}$ is the array of costs for each node.

\textit{Different stooge behavior}: In the pseudocode we only change resistances, but we can modify the algorithm to only change opinions, or even change both opinions and resistances. If we only allow opinions to change, then when a node becomes a stooge we just need to recalculate its value based on the fact that his opinion is 1.0. If we can only modify the resistances, then when we make a node a stooge, he will either set the resistance to be 0 or 1. This is optimal in this case since we want to maximize the node's opinion, and this opinion will be a linear combination of the stooge's opinions and his opinion. Because of that the max will be achieved at one of the extreme points. Finally if you can change both opinion and resistance, then the dp state of that node that becomes stooge is $1$, since we will set opinion $1$ for the node, and make resistance also be $1$, thus he will return $1$ if any random walk enters this node.

Finally we are also allowed to have self loops. If we have that, we just look at the edge's weight and multiply by $\frac{1}{\text{edgeWeight}}$ to get the final opinion of a node.

\subsection{Dynamic programming experiments}

\begin{figure}[ht]
  \centering
  \includegraphics[width=0.32\linewidth]{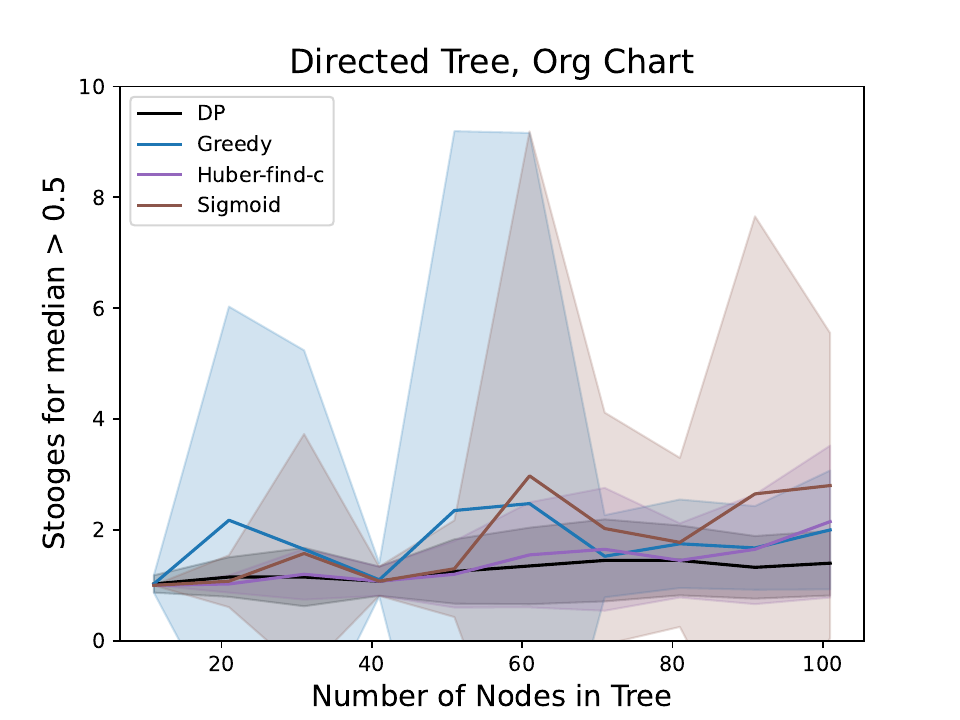}
  \includegraphics[width=0.32\linewidth]{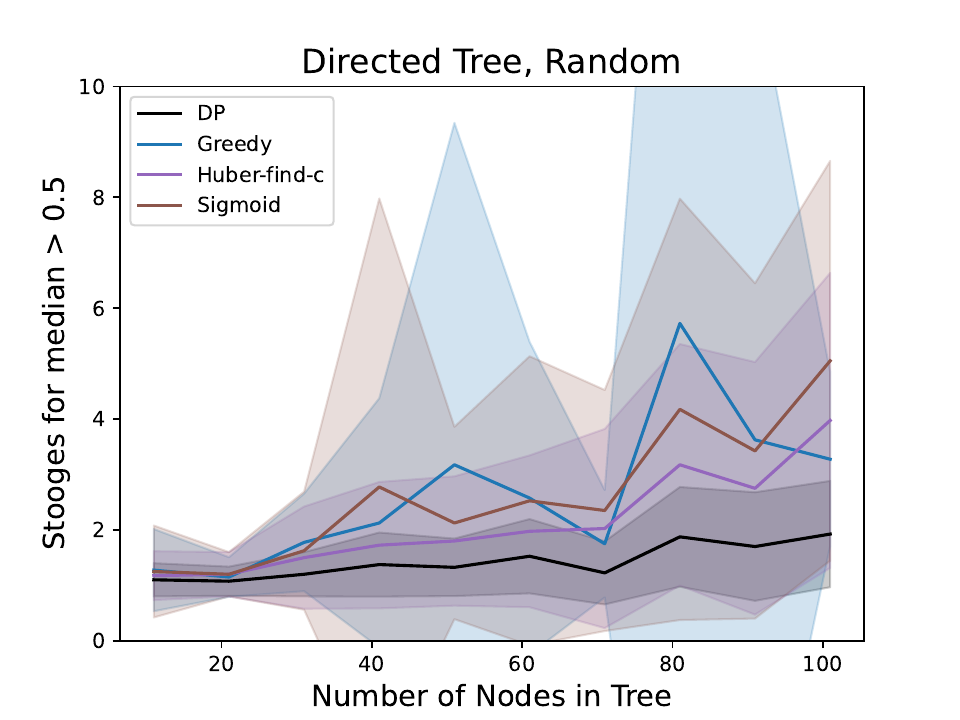}
  \includegraphics[width=0.32\linewidth]{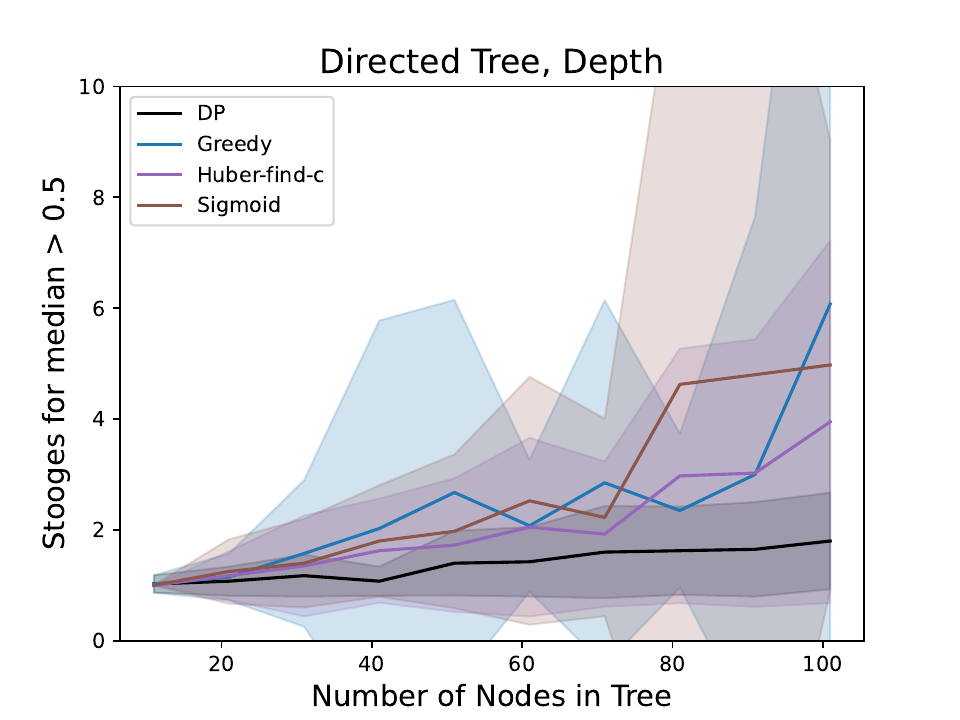}
  \caption{Comparison of DP vs Greedy Approaches: (Left) Simulated Org Chart, (Center) Random Tree, (Right) Depth Tree}
  \label{fig:dyn-comparison}
\end{figure}

\begin{figure}[ht]
  \centering
  \includegraphics[width=0.32\linewidth]{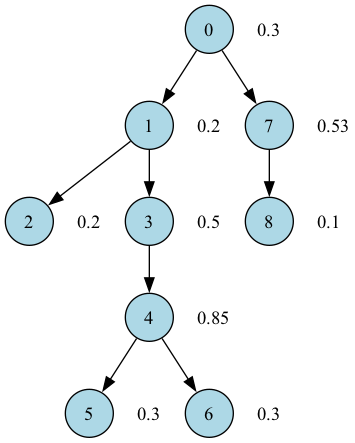}
  \includegraphics[width=0.32\linewidth]{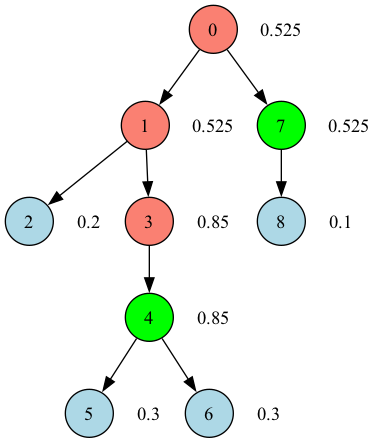}  \includegraphics[width=0.32\linewidth]{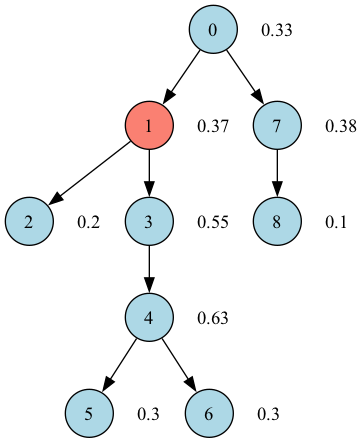}
  \caption{A difficult instance. On the left, we show the tree
 before introducing stooges. In the middle, we show the tree with optimally assigned stooge. On the right, we show the outcome of our greedy algorithm. Nodes with resistance $1$ are green, nodes with resistance $0$ are red.}
  \label{fig:dyn-hard}
\end{figure}

In order to compare the 4 algorithms (Dynamic-Programming, Greedy, Huber-find-c, and Sigmoid), we run them on randomly generated trees until all of them find a solution. It is expected for the non-optimal algorithms to fail to find a solution sometimes, mostly due to how restrictive the graph type is. Failure to find a solution is less common in natural graphs, or even trees where the information flows differently.
We run the algorithms on the following on random trees, organizational chart type trees (every node has between 2-5 sub-ordinates as in a real company), and  depth trees, where we attach nodes in sequence. Here, every new node is attached to one of the 5 previously added nodes. This guarantees a depth of at least $\frac{N}{5}$.

\subsubsection{Difficult instances for non-dynamic programming solutions}

In Figure~\ref{fig:dyn-hard} we can see an instance where all algorithms other than the dynamic programming fail. This is because the setup has only one solution, so any mislabeling of a non-leaf node results in failure to find a solution. Additionally, some nodes seem to exhibit unintuitive behavior (such as node 3 becoming a stooge while already having an opinion above 0.5). In order to properly compare the performance of our algorithms, we only looked at graphs where all of them succeed. One thing to note is that these hierarchical graphs are the only type of graphs where we found it hard for all algorithms to find solutions. 

The greedy only finds one node with marginal gains, then stops since it cannot improve the median further since all changes lead to the same median.

\section{Hardness}

\textsc{Theorem~\ref{thm:hardness}}.
\emph{Problem~\ref{prob:max-median}
for $p=0$
is inapproximable
to any multiplicative factor
unless $\mathsf{P} = \mathsf{NP}$.}

\begin{proof}
We use a reduction from set cover
~\cite{williamsonbook}.
Let $U = \{1, 2, \dots, n\}$,
subsets $S_1, \dots, S_m \subseteq U$,
and a budget $k$ be
an instance for set cover.
We create an instance for
the generalized FJ dynamics.
We use the following graph
that has a vertex $u_i$ for every
element $1 \le i \le n$,
and the two vertices $v_j$ and $w_j$
for every set $1 \le j \le m$.
We use innate opinions $s(u_i) = 0$
and resistances $\alpha(u_i) = 0$
for all $1 \le i \le n$. For the
remaining vertices, we set
$s(v_j) = 0$ and $s(w_j) = 1$
as well as $\alpha(v_j) = 1$
and $\alpha(w_j) = 1$
for all $1 \le j \le m$.
We add a directed edge $(u_i, v_j)$
whenever $i \in S_j$.
We also add a directed edge $(v_j, w_j)$
for all $1 \le j \le m$.
Finally, we introduce $\ell = n + 2k$
isolated nodes with opinion $0$ such
that there is a total of $\ell+n+2m = 2(n+m+k)$
nodes in the constructed network.
The motivation behind this construction
is due to the absorbing random walk
interpretation for opinion formation
~\cite{gionis2013opinion}. Here, we
simulate a random walk starting from
any node $u$ in the graph.
In each step of the random walk,
we are absorbed with probability
$\alpha(v)$ where $v$ is the vertex
the random walk currently visits. If the
random walk gets absorbed, we assume
the innate opinion $s(v)$. It can be
shown that in expectation, the opinion
$x^\star(u)$ is the
equilibrium opinion.
In our construction, random walks
from the vertices $u_i$ for $1 \le i \le n$
are initially all absorbed by
vertices $v_j$ and therefore have
opinion $x^\star_{v_j} = 0$. Now,
we want that selecting a
set $1 \le j \le m$ corresponds to changing
the resistance of $v_j$ to $\alpha'(v_j) = 0$.
As such, whenever an element $u_i$ is
covered by a selected set $S_j$, there is a
non-zero probability that the random
walk will go over $v_j$ to $w_j$ where
it assumes opinion $s(w_j) = 1$.
As such, the opinion of covered elements
changes to $>0$.
We show that there is a solution
to the set cover instances if
and only if the $\mathrm{Median}(x^\star) > 0$.

$(\Rightarrow)$ Assume now there
is a solution $J \subseteq \{1, \dots, m\}$
to the set cover instance
with $|J| \le k$ that covers all elements.
In this case, we can choose
$\alpha'$ which is equal to $\alpha$
except on $v_j$ for all $j \in J$
where we set $\alpha'(v_j) = 0$.
By our construction, a random walk
starting from every vertex $u_i$ has
non-zero probability to move to
a node $v_j$ for $j \in J$ such
that $u_i \in S_j$. Since we defined
$\alpha'$ such that $v_j$ is not
absorbing, the random walk continues
to $w_j$ where it is absorbed.
As such, the opinion of all nodes
$u_i$ is $x^\star(u_i) > 0$.
In total, there are $n+k+m$ nodes
with opinion $>0$, which is
half the total number of nodes.
In this case, we
break ties such that the median
becomes $\mathrm{Median}(x^\star) >0$.
Note that this assumption is
only for simplicity, as we could
achieve the same by, for instance,
duplicating all nodes $u_i$.

$(\Leftarrow)$ Assume there is no
solution to the set cover instance.
Note that modifying either the
resistances of vertices $u_i$ or $w_j$
cannot increase the opinion of any node.
Since we want to increase opinion values,
we therefore choose without loss
of generality modified resistances $\alpha'$
where only $k$ resistance values on nodes
$v_j$ differ. However, since no set
of $k$ sets covers the universe, there is
at least one node $u_i$ with opinion
$x^\star(u_i) = 0$.
In total, there are at most $n-1+k+m$
nodes with opinion $>0$.
Since this is less than half of the
nodes, we have $\mathrm{Median}(x^\star) = 0$.

As such, any (multiplicative) approximation
algorithm will be able to differentiate
the two cases. The existence of any
such algorithm would therefore imply that
$\mathsf{P} = \mathsf{NP}$.
\end{proof}

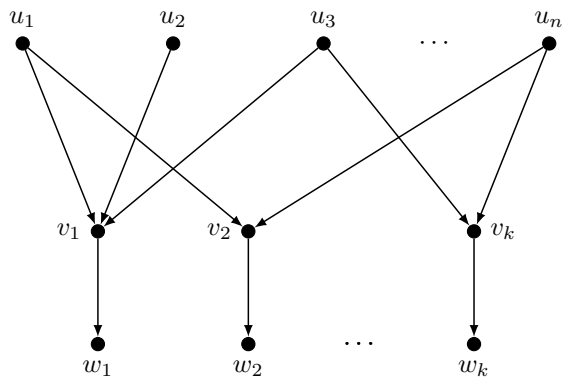
\begin{figure}
\centering
\begin{tikzpicture}
    \tikzstyle{vertex}=[draw, circle, fill=black, inner sep=0pt, minimum size=5pt]

    \node[vertex, label=above:$u_1$] (u1) at (0, 0) {};
    \node[vertex, label=above:$u_2$] (u2) at (2, 0) {};
    \node[vertex, label=above:$u_3$] (u3) at (4, 0) {};
    \node at (5.5, 0) {$\cdots$};
    \node[vertex, label=above:$u_n$] (un) at (7, 0) {};

    \node[vertex, label=left:$v_1$] (v1) at (1, -2.5) {};
    \node[vertex, label=left:$v_2$] (v2) at (3, -2.5) {};
    \node[] (ss) at (4.5, -4) {$\cdots$};
    \node[vertex, label=right:$v_k$] (vm) at (6, -2.5) {};

    \node[vertex, label=below:$w_1$] (w1) at (1, -4) {};
    \node[vertex, label=below:$w_2$] (w2) at (3, -4) {};
    \node[] (ss) at (4.5, -4) {$\cdots$};
    \node[vertex, label=below:$w_k$] (wm) at (6, -4) {};

    \draw[-latex,line width=0.2mm] (u1) to (v1);
    \draw[-latex,line width=0.2mm] (u2) to (v1);
    \draw[-latex,line width=0.2mm] (u3) to (v1);
    \draw[-latex,line width=0.2mm] (u1) to (v2);
    \draw[-latex,line width=0.2mm] (un) to (v2);
    \draw[-latex,line width=0.2mm] (u3) to (vm);
    \draw[-latex,line width=0.2mm] (un) to (vm);

    \draw[-latex,line width=0.2mm] (v1) to (w1);
    \draw[-latex,line width=0.2mm] (v2) to (w2);
    \draw[-latex,line width=0.2mm] (vm) to (wm);
\end{tikzpicture}
\caption{Gadget used in the proof of
Theorem~\ref{thm:hardness}. We reduce from a set cover
instance with $U = \{1, 2, \dots, n\}$ and subsets
$S_1 = \{1, 2, 3\}, S_2 = \{u_1, u_n\}, \dots,
S_k = \{u_3, u_n\}$.}
\end{figure}

\textsc{Corollary~\ref{cor:quantile}.}
\emph{
Let $(\alpha, W, s)$ be a
network of generalized FJ  dynamics,
let $k$ be a budget and $0 < q < 1$
be any quantile.
The problem to maximize the $q$-th
quantile of $x^\star(\alpha', W, s)$
subject to $\|\alpha' - \alpha\|_0 \le k$
cannot be approximated
to any multiplicative factor
unless $\mathsf{P} = \mathsf{NP}$.
}

\begin{proof}
This follows through a simple adaption
of the proof of Theorem~\ref{thm:hardness}.
Specifically, we can change the point
where the $q$-th quantile becomes positive
by modifying the number of isolated vertices to
$\ell=\lfloor (\frac 1 q - 1) n + (\frac 1 q - 2) m + \frac 1 q k \rfloor$.
This ensures
that the $q$-th percentile is $>0$
if and only if there exists a set cover.
\end{proof}

\end{document}